\documentclass[12pt]{article}
\usepackage{setspace}

\usepackage{amsmath, graphics, graphicx}
\usepackage{amsthm}
\usepackage{amsfonts}
\usepackage{natbib}
\usepackage{float}
\usepackage{booktabs}
\usepackage{multirow}
\usepackage{appendix}

\usepackage{tabularx} 

\newtheorem{assumption}{Assumption}

\usepackage[colorlinks = true, linkcolor=blue, citecolor = blue]{hyperref}

\usepackage[normalem]{ulem} 

\newcommand{\Robs}{\ensuremath{R^{\rm obs}}}
\newcommand{\robs}{\ensuremath{r^{\rm obs}}}

\newcommand{\na}{\textsf{NA}}
\newcommand{\gateaux}{\textsf{Gateaux}}

\newcommand{\surv}{\mbox{surv}}

\newcommand{\infoE}{Z}
\newcommand{\infoL}{W}
\newcommand{\data}{O}
\newcommand{\mas}{\Omega}

\newcommand{\ci}{\perp}
\newcommand{\pr}{\ensuremath{\mbox{pr}}}

\newcommand{\cL}{\mathcal{L}}
\newcommand{\cF}{\mathcal{F}}

\usepackage{dsfont}
\newcommand{\indic}{{\mathds 1}}

\usepackage[usenames, dvipsnames]{color}
\definecolor{mypurple}{RGB}{203, 66, 244}
\definecolor{mygray}{RGB}{145,145,145}
\definecolor{mygreen}{RGB}{0,128,0}
\definecolor{myorange}{RGB}{255,99,71}

\begin{document}

\title{Deductive semiparametric estimation in Double-Sampling Designs with application to PEPFAR}

\author{Tianchen Qian\thanks{Department of Statistics, Harvard University. qiantianchen@fas.harvard.edu} 
\and Constantine Frangakis\thanks{Department of Biostatistics, Johns Hopkins University.}
\and Constantin Yiannoutsos\thanks{Department of Biostatistics, Indiana University.}}



\date{\today}

\maketitle

\begin{abstract}

Non-ignorable dropout is common in studies with long follow-up time, and it can bias study results unless handled carefully in the study design and the statistical analysis. A double-sampling design allocates additional resources to pursue a subsample of the dropouts and find out their outcomes, which can address potential biases due to non-ignorable dropout. It is desirable to construct semiparametric estimators for the double-sampling design because of their robustness properties. However, obtaining such semiparametric estimators remains a challenge due to the requirement of the analytic form of the efficient influence function (EIF), the derivation of which can be \textit{ad hoc} and difficult for the double-sampling design. Recent work has shown how the derivation of EIF can be made deductive and computerizable using the functional derivative representation of the EIF in nonparametric models. This approach, however, requires deriving the mixture of a continuous distribution and a point mass, which can itself be challenging for complicated problems such as the double-sampling design. We propose semiparametric estimators for the survival probability in double-sampling designs by generalizing the deductive and computerizable estimation approach. In particular, we propose to build the semiparametric estimators based on a \textit{discretized support} structure, which approximates the possibly continuous observed data distribution and circumvents the derivation of the mixture distribution. Our approach is \textit{deductive} in the sense that it is expected to produce semiparametric locally efficient estimators within finite steps without knowledge of the EIF. We apply the proposed estimators to estimating the mortality rate in a double-sampling design component of the President's Emergency Plan for AIDS Relief (PEPFAR) program. We evaluate the impact of double-sampling selection criteria on the mortality rate estimates. Simulation studies are conducted to evaluate the robustness of the proposed estimators.

\end{abstract}

\noindent%

\maketitle

\newpage

\onehalfspacing

\section{Introduction}
\label{sec:intro}

Studies with long follow-up often suffer from a high dropout rate. Dropouts can depend on the outcome of interest, even after adjusting for observed covariates. This makes the dropouts ``non-ignorable'' and biases the analysis based solely on the non-dropouts \citep{Rubin76}. As a way to handle non-ignorable dropouts, double-sampling designs allocate additional resources to pursue a sample of the dropouts and find out their outcomes \citep{baker1993,glynn1993,frangakis2001,cochran2007sampling}.
Compared to standard double-sampling (where each dropout is sampled with equal probability), the double-sampling can be more practical or informative if it targets dropouts whose history at the dropout time has specific profiles.  For example, \citet{an2009biometrics} found that such ``profile designs'' can save more than 35\% of resources compared to the standard double-sampling design. It can also be more practical to double-sample relatively recent dropouts. For data analysis of such profile double-sampling designs, \citet{an2014statmed} employed a parametric approach to estimate the survival probability. However, analyses of such designs can be more reliable if they do not rely heavily on parametric assumptions. Semiparametric estimators \citep{newey1990semiparametric, tsiatis07}, which focus on modeling the parameter of interest and treat the rest of the model as nuisance parameters, have had great success in various areas \citep{cox1972regression, liang1986longitudinal, robins1994estimation, zhang2008improving}, and they are a promising alternative to the parametric estimators for double-sampling designs. \citet{robins2001discussion} had suggested a possible way of deriving a semiparametric estimator for double-sampling designs, but that and any other such existing proposals rely on first coming up with and then verifying conjectures ``by hand''. Such a process is prone to human errors and is not generalizable.


Recently, \citet{frangakis2015deductive} proposed a different approach to construct semiparametric estimators. Contrary to the classical semiparametric framework \citep{bickel1993efficient, tsiatis07} which relies heavily on the conjecture and verification of the analytic form of the efficient influence function (EIF), their approach produces semiparametric locally efficient estimators by utilizing the Gateaux derivative representation of the EIF in nonparametric models. This estimation procedure is \textit{deductive}, in the sense that the estimator is computed through a discrete and finite set of steps, without requiring conjecturing for or theoretically deriving the EIF. If a semiparametric estimator can be obtained deductively, human effort can be saved from difficult mathematical derivations and can be transferred, for example, to designing new studies. Their approach is not directly applicable to the data analysis of double-sampling designs, however, because their estimation procedure requires analytically evaluating the parameter of interest at a mixture of a continuous distribution and a point mass. This analytic step is feasible for certain problems (e.g., for the two-phase design considered in \citet{frangakis2015deductive}), but can become highly error-prone when derived by hand in complicated problems like estimation of the survival probability in the double-sampling design.

In this paper, we develop semiparametric estimators for the survival probability in double-sampling designs, by generalizing the deductive estimator by \citet{frangakis2015deductive} to problems where the aforementioned analytic evaluation is infeasible. We use a \textit{discretized support} structure to circumvent the analytical evaluation of the parameter of interest at a mixture distribution. The discretized support is a discretized approximation of the sample space constructed from the observed data, which enables numerical calculation of the estimand at the mixture. A similar discretization idea was used in \citet{chamberlain1987} for deriving efficiency bounds in nonparametric models. Our method produces semiparametric locally efficient estimators within finite steps without knowledge of the EIF; i.e., it is deductive.

We apply the method to estimating the mortality rate using data from a double-sampling design component of the President's Emergency Plan for AIDS Relief (PEPFAR), an HIV monitoring and treatment program \citep{geng2015ART}. In addition, we explore and discuss how certain restrictions on the double-sampling selection criteria may impact the estimated mortality rate, as the double-sampling for this study is more pragmatic if restricted to relatively recent dropouts \citep{an2014statmed}.

The paper is organized as follows. In Section \ref{sec:design}, we introduce the double-sampling design and the parameter of interest. In Section \ref{sec:method}, we present the proposed estimation method for survival probability in double-sampling designs. In particular, we present and discuss how the discretized support idea facilitates the deductive estimation in double-sampling designs. In Section \ref{sec:simulation}, we present results from simulation studies which show the performance of the proposed estimators and several other existing methods. In Section \ref{sec:application}, we apply the proposed method to estimating the mortality rate using data from a double-sampling design component of PEPFAR, and we discuss the impact of restrictions on the double-sampling selection criteria on the estimates. Section \ref{sec:discussion} concludes with discussion.

\section{Double-sampling design and parameter of interest}
\label{sec:design}

\subsection{Double-sampling design}

\begin{figure}[htbp]
\begin{center}
\includegraphics[width = \textwidth]{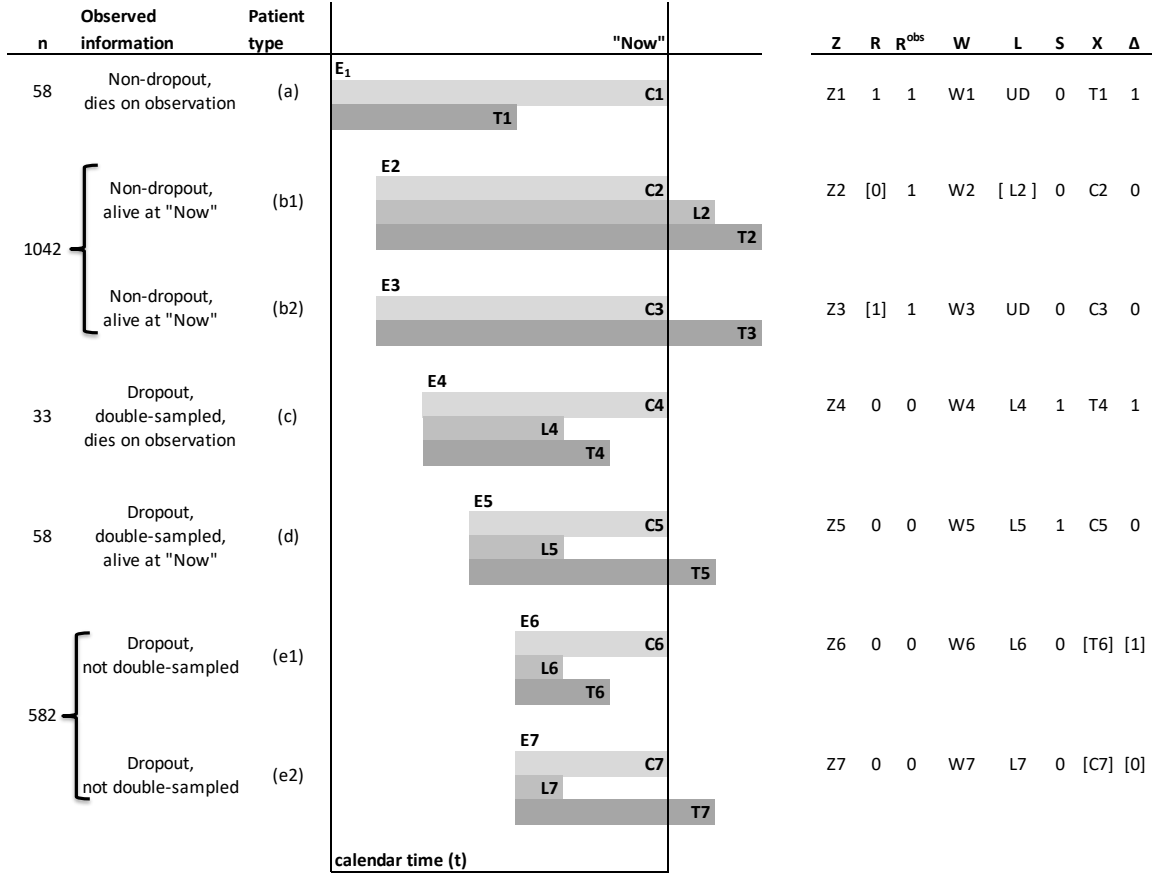}
\vspace{-5.5cm}
\caption{Characteristics for different patient types in a double-sampling design. $E_i$ denotes the enrollment time; $C_i$ denotes the time to administrative censoring; $T_i$ denotes the time to death; $L_i$ denotes the time to dropout. $Z$ denotes covariates observed at enrollment; $R$ denotes the true dropout status; $\Robs$ denotes the observed dropout status; $W$ denotes the dropout time and some other longitudinal measure; $S$ denotes the double-sampling selection indicator; $(X, \Delta)$ is the standard survival outcome: $X = \min(T,C)$, $\Delta = \indic(T \leq C)$, where $\indic(\cdot)$ is the indicator function. ``UD'' means undefined. Quantities in brackets are unobserved. Note that unobserved $L$ and undefined $L$ cannot be distinguished from the observed data. Column $n$ denotes the number of patients of each type in the double-sampling data set of PEPFAR. The figure is a modified version of Figure 1 in \citet{an2014statmed}.}
\label{fig:design}
\end{center}
\end{figure}

For clarity, we present arguments with a double-sampling design as shown in Figure \ref{fig:design}, which illustrates different patient types in the double-sampling component of PEPFAR. This figure is a modified version of Figure 1 in \citet{an2014statmed}.
First, we describe characteristics that are inherent to a patient (i.e., potential outcomes \citep{Rubin76}), which would be realized if the program could enroll and follow a patient indefinitely with a standard effort. Then, we describe the actual double-sampling design consisting of two phases and the observed data.

\bigskip

\noindent \textit{Patient inherent characteristics and parameter of interest.} For each patient, there is an enrollment time $E_i$, covariates $\infoE_i$ at enrollment, and a survival time $T_i$ (time from enrollment to death). The parameter of interest is $P(T_i > t)$ for a given $t$, which is the proportion of patients surviving beyond year $t$ for a population represented as a random sample by the enrolled patients. If the program were to follow patients indefinitely with a standard effort while they are alive, some patients would discontinue contact from the monitoring (patient types (b1), (c), (d), (e1), and (e2) in Figure \ref{fig:design}). For these patients, labeled true dropouts and indicated by $R_i=0$, denote by $L_i$ the time from enrollment to dropout and denote by $\infoL_i$ any information available at dropping out in addition to $\infoE_i$. For instance, $\infoL_i$ may include $L_i$ and some longitudinal measurements.

\bigskip

\noindent \textit{Phase 1 of the double-sampling design and missingness due to administrative censoring.} In the first phase, the actual design enrolls patients at times $E_i$ and monitors them with standard effort, not indefinitely, but until ``Now''---the present time. The time from enrollment $E_i$ to ``Now'' is the time to administrative censoring, denoted by $C_i$. For patient $i$, define $\Delta_i=1$ if $C_i\geq T_i$ and $\Delta_i=0$ otherwise. Define $X_i=\min(C_i,T_i)$.

Simply based on Phase 1, the standard survival data $(X_i, \Delta_i)$ are not observed for dropouts whose time to dropout satisfies $L_i < X_i$. Denote such \textit{observed} dropout patients by $\Robs_i=0$ (patient types (c), (d), (e1), and (e2) in Figure \ref{fig:design}), and denote the rest (\textit{observed} non-dropouts) by $\Robs_i=1$.

\bigskip

\noindent \textit{Phase 2: information after double-sampling.} Phase 2 of the design selects a subset of the observed dropouts, called a double-sample, and allocates additional resources for searching for and finding this double-sample at ``Now''. Denote by $S_i$ the indicator of being selected as a double-sample. The survival data $(X_i, \Delta_i)$ is observed for the double-samples (those with $S_i = 1$). 

The observed data for patient $i$ is $(C_i, \Robs_i, Z_i, W_i, S_i, X_i, \Delta_i)$. For the patients with $\Robs_i = 1$, we define $S_i = 0$. For the patients with $\Robs_i = 0$ and $S_i = 0$, the survival data $(X_i,\Delta_i)$ is not observed, and we define $(X_i,\Delta_i) = (\na, \na)$. If $W_i$ only includes the dropout time $L_i$, $W_i$ is undefined for the patients with $\Robs_i = 1$ and we denote it by $\na$; if $W_i$ includes additional longitudinal measurement that is observed for those $\Robs_i = 1$, then $W_i$ is defined for all patients. We use $\perp$ to denote independence between random variables.

\subsection{Identification of parameter}

To identify the parameter of interest $P(T_i > t)$ by the observed data distributions, we make the following assumptions.

\begin{assumption}[Patient-dependent double-sampling] \label{assumpt:double-sampling}
For each observed dropout, $S_i$ can depend on the patient characteristics $(\infoE_i,\infoL_i)$; after we condition on $(\infoE_i, \infoL_i)$, double-sampling $S_i$ is independent of survival and enrollment times:
\begin{equation*}
    S_i \ci (T_i,E_i) \mid \Robs_i=0, \infoE_i,  \infoL_i.
\end{equation*}
\end{assumption}

\begin{assumption}[Exchangeability of enrollment times] \label{assumpt:independent-censoring}
Patients enrolled at different times (or equivalently, patients having different time to administrative censoring $C_i=\mbox{``Now''}-E_i$) have similar survival times after conditioning on $\infoE_i$: 
\begin{equation*}
    T_i \ci C_i \mid \infoE_i .
\end{equation*}
\end{assumption}

Assumption \ref{assumpt:double-sampling} means that the probability of double-sampling an observed dropout can depend on his/her characteristics known up to the dropout time $L_i$, but not depending on other unobserved information (such as prognostic information for the survival that are not included in $(Z_i, W_i)$). This can be ensured by including in $(Z_i, W_i)$ the variables used by the investigators to select the double-samples. Assumption \ref{assumpt:independent-censoring} means that there are no time-trend in the survival times after conditional on $Z_i$; this is plausible in practice when the design is limited to a window period of enrollment. Intuitively, Assumption \ref{assumpt:double-sampling} helps to recover the information missing due to dropout, and Assumption \ref{assumpt:independent-censoring} helps to recover the information missing due to administrative censoring. 

Throughout the paper, we use $P(\cdot)$ to denote the probability of an event or a point in the sample space, and we use $\pr(\cdot)$ to denote the probability density function of a random variable with with respect to an appropriate dominating measure (the Lebesgue measure for continuous variables and the counting measure for discrete variables). Under Assumptions \ref{assumpt:double-sampling} and \ref{assumpt:independent-censoring}, the estimand $P(T_i > t)$ is identifiable from the following components of the observed data distribution: 
\begin{align}
& P(\Robs_i = 1); && \pr(\infoE_i \mid \Robs_i=1); && \pr\{(X_i,\Delta_i)\mid \infoE_i, \Robs_i=1\}; \label{eq:distr1}\\
& P(\Robs_i = 0); && \pr(\infoE_i,\infoL_i \mid \Robs_i=0); && \pr\{(X_i,\Delta_i)\mid \infoE_i,\infoL_i, \Robs_i=0\}, \label{eq:distr2}
\end{align} 
In particular, by Assumption \ref{assumpt:double-sampling} the distribution from the double-sampled individuals $\pr\{(X_i,\Delta_i)\mid \infoE_i,\infoL_i, \Robs_i=0, S_i=1\}$ is the same as the one for all dropouts, $\pr\{(X_i,\Delta_i)\mid \infoE_i,\infoL_i,$ $\Robs_i=0\}$, and so, together with the second component of \eqref{eq:distr2} gives, upon averaging over $ \infoL_i$,
 $\pr\{(X_i,\Delta_i),\infoE_i \mid \Robs_i=0\}$.
That, together with \eqref{eq:distr1}, gives $\pr\{(X_i,\Delta_i),\infoE_i\}$. Denote by $\surv_t(\cdot)$ the function that takes $\cdot$ as an arbitrary distribution of $(X_i,\Delta_i)$ from independent survival and censoring times and returns the survival probability beyond $t$ (this function is the common probability limit of the Nelson-Aalen and Kaplan-Meier estimators). Then, by Assumption \ref{assumpt:independent-censoring}, the estimand can be expressed in terms of the observed data distribution as
\begin{equation}
\tau = P(T_i > t) = \int \surv_t\left[\pr\{(X_i,\Delta_i)\mid \infoE_i\} \right ] \pr(\infoE_i) d(\infoE_i). \label{eq:estimand1}
\end{equation}
A detailed derivation of \eqref{eq:estimand1} is given in Appendix \ref{appen:identification}.

\section{Method}
\label{sec:method}

\subsection{Review of the estimation procedure by \citet{frangakis2015deductive}}
\label{subsec:review-deductive-estimator}

An estimand can be viewed as a functional of a distribution on the data. For example, in the double-sampling design described above, \eqref{eq:estimand1} shows that $P(T_i > t)$ is a functional $\tau(F)$ of the distribution $F$ of the observed data. Estimation of $\tau(F)$ requires modeling assumptions because \eqref{eq:estimand1} involves regressions that, in practice, cannot be estimated fully nonparametrically. In particular, suppose the estimand $\tau$ in \eqref{eq:estimand1} has a nonparametric efficient influence function (EIF) denoted by $\phi(\data_i,[F-\tau,\tau])$, where $O_i$ represents the observed data from subject $i$ and $F-\tau$ represents the remaining components of the distribution other than $\tau$. A semiparametric estimator $\hat\tau$ can be obtained by solving 
\begin{equation}
\sum_i \phi\{\data_i,[(F-\tau),\tau]\} = 0 \label{eq:estimator}
\end{equation}
after substituting for $(F-\tau)$ estimates of a working model. Such an estimator is consistent and locally efficient if the working estimators of ($F-\tau)$ are consistent with convergence rates larger than $n^{1/4}$ \citep{van2000asymptotic}. In classical semiparametric theory \citep{bickel1993efficient, tsiatis07}, obtaining such a semiparametric estimator requires knowing the analytic form of the EIF $\phi$, the derivation of which usually relies on first conjecturing for and then verifying the form of certain Hilbert spaces. Although this process has been successful in some settings \citep{robins1994estimation, zhang2008improving}, it is not guaranteed to succeed for estimation in double-sampling designs.

An alternative approach to construct semiparametric estimators was proposed by \citet{frangakis2015deductive}, which does not require the analytic form of $\phi$. The key idea in \citet{frangakis2015deductive} is that, for any working distribution $F(\alpha)$ parametrized by $\alpha$, one can approximate numerically the EIF $\phi$ as the numerical Gateaux derivative after perturbing the working distribution $F(\alpha)$ by a small mass at each observed data point. Formally, let $\cL_{O_i}$ be a point mass at $O_i$, and let $F_{(\data_i,\epsilon)}(\alpha)=(1-\epsilon) F(\alpha) + \epsilon \cL_{O_i}$ be the perturbed distribution, i.e., a mixture of $F(\alpha)$ and $\cL_{O_i}$, where $\epsilon>0$ is a small positive number. One can approximate $\phi\{\data_i,  F(\alpha)\}$ numerically by $\gateaux\{\data_i,F(\alpha),\epsilon\}$, where
\begin{equation}
    \gateaux\{\data_i,F(\alpha),\epsilon \}:= \left [\tau\{F_{(\data_i,\epsilon)}(\alpha)\}- \tau\{F(\alpha)\}\right ]/{\epsilon}. \nonumber
\end{equation}
Then, assuming $\alpha$ is 1-dimensional, one can find the best working distribution parameter $\hat{\alpha}$ as one that makes zero the sum of the numerical EIFs, $\sum_i \gateaux\{\data_i,F(\alpha),\epsilon \}$. The estimator solving \eqref{eq:estimator} approximately is then $\tau\{ F(\hat{\alpha})\}$. The standard error of the estimator can be estimated by $n^{-1}\big[ \sum_i \gateaux\{\data_i,F(\hat{\alpha}),\epsilon \}^2 \big]^{1/2}$ with $n$ denoting the sample size. 

The estimator $\tau\{ F(\hat{\alpha})\}$ has consistency properties beyond those of the maximum likelihood estimator for the same working model $F(\alpha)$, since the former depends only on the parts of $F$ that are present in the nonparametric influence function EIF. For example, \cite{frangakis2015deductive} show that in the two-phase design, $\tau\{ F(\hat{\alpha})\}$ shares the double robustness of the estimators require the form of the EIF. A general condition regarding the true distribution $F_0$ and the working model $F$ for the consistency of $\tau\{ F(\hat{\alpha})\}$ is stated in Appendix \ref{appen:robust}.


\subsection{Proposed method: deductive estimation with discretized support}
\label{subsec:algorithm}


A semiparametric estimator is \textit{deductive}, if it does not rely on conjectures for the functional form of the EIF $\phi$ and it is produced in the sense of \cite{turing} (i.e., using a discrete and finite set of instructions, and, for every input, finishing in discrete finite steps). $\tau\{ F(\hat{\alpha})\}$ in Section \ref{subsec:review-deductive-estimator} is such an example. Deductive estimators are desired for analysis of the double-sampling designs, because they often have additional robustness properties that the maximum likelihood estimator does not have, and they do not require the difficult derivation of the EIF. However, directly application of the deductive estimation procedure in \citet{frangakis2015deductive} to the double-sampling design is infeasible, because of the difficulty in evaluating $\tau\{F_{(\data_i,\epsilon)}(\alpha)\}$.

We generalize the deductive estimation procedure to double-sampling designs by incorporating the so-called \textit{discretized support}. The purpose of the discretized support is to approximate possibly continuous working models (i.e., $F(\alpha)$) via discretization. Intuitively, if the support of a discretized working model contains all the observed data points, perturbing the working model by a point mass at an observed data point (i.e., $F_{(\data_i,\epsilon)}(\alpha)$) would be as simple as changing the values in a probability table, and this would circumvent the analytic evaluation of $\tau\{F_{(\data_i,\epsilon)}(\alpha)\}$. A similar discretization idea was used in \citet{chamberlain1987} for deriving nonparametric efficiency bounds.

In the following we use lower case letter to denote the realized value of random variables. Recall that the observed data from patient $i$ in the double-sampling design is $(c_i, \robs_i, z_i, w_i, s_i, x_i, \delta_i)$, and we use $O_i$ to denote the observed data from patient $i$ excluding $c_i$; i.e., $O_i = (\robs_i, z_i, w_i, s_i, x_i, \delta_i)$.
Suppose that among the $n$ total patients there are $m$ observed dropouts ($\robs_i = 0$) and $m_1$ patients get double-sampled ($\robs_i = 0, s_i = 1$), and that the data is ordered such that $\robs_1 = \robs_2 = \cdots = \robs_m = 0$, $\robs_{m+1} = \cdots = \robs_n = 1$, $s_1 = s_2 = \cdots = s_{m_1} = 1$, and $s_{m_1 + 1} = \cdots = s_n = 0$. The discretized support for those with $\robs_i = 0$ is defined as
\begin{align*}
    \mas_0 = \{0\} \otimes \{ (z_1, w_1), \ldots, (z_m, w_m) \} \otimes \{ (s_1, x_1, \delta_1), \ldots, (s_m, x_m, \delta_m) \},
\end{align*}
where $\otimes$ denotes Cartesian product\footnote{The set operation $\{\cdot\}$ removes duplicates. In particular, $\{ (z_1, w_1), \ldots, (z_m, w_m) \}$ is the set of all unique pairs $(z_i, w_i)$ for $1\leq i \leq m$, and $\{ (s_1, x_1, \delta_1), \ldots, (s_m, x_m, \delta_m) \}$ is the set of all unique triples $(s_i, x_i, \delta_i)$ for $1\leq i \leq m$.}. Each element in $\mas_0$ is a vector of the same length as $(\Robs, Z, W, S, X, \Delta)$, and we have $O_i \in \mas_0$ for $1 \leq i \leq m$.
The discretized support for those with $\robs_i = 1$ is defined as
\begin{align*}
    \mas_1 = \{1\} \otimes \{ (z_{m+1}, w_{m+1}), \ldots, (z_n, w_n) \} \otimes \{ (s_{m+1}, x_{m+1}, \delta_{m+1}), \ldots, (s_n, x_n, \delta_n) \},
\end{align*}
and similarly we have $O_i \in \mas_1$ for $m+1 \leq i \leq n$.
The overall discretized support is defined as
\begin{align}
    \mas = \mas_0 \cup \mas_1, \nonumber
\end{align}
and we have $O_i \in \mas$ for $1 \leq i \leq n$.

The deductive estimation procedure in \citet{frangakis2015deductive} is now feasible for the double-sampling design by conducting all the computations on the discretized support $\Omega$. Suppose we have a class of fitted working distributions $\hat{F}(\alpha)$ on $\Omega$, parametrized by a 1-dimensional $\alpha$. In other words, for each value of $\alpha$, $\hat{F}(\alpha)$ is a discrete probability distribution on the sample space $\Omega$. The particular form of $\hat{F}(\alpha)$ will be described in detail in Section \ref{subsec:F-alpha}. Because $O_i \in \Omega$ for all $i$, the mixture distribution $\hat{F}_{(\data_i,\epsilon)}(\alpha) = (1-\epsilon) \hat{F}(\alpha) + \epsilon \cL_{\data_i}$ with a small $\epsilon>0$ is also a discrete probability distribution on $\Omega$, and can be computed directly from $\hat{F}(\alpha)$ and $O_i$.

For any probability distribution $G$ on $\Omega$, it is straightforward to compute the induced distribution components \eqref{eq:distr1} and \eqref{eq:distr2} through averaging and normalization; details are provided in Appendix \ref{appen:averaging-normalization} for completeness. Therefore, the estimand $\tau(G)$ can be computed by \eqref{eq:estimand1} nonparametrically because the sample space of $G$ is discrete. We compute $\tau\{\hat{F}(\alpha)\}$ and $\tau\{\hat{F}_{(\data_i,\epsilon)}(\alpha)\}$ for all $1\leq i\leq n$, and find $\hat\alpha$ that solves
\begin{align}
    \sum_{i=1}^n \gateaux\{O_i, \hat{F}(\alpha), \epsilon\} = \sum_{i=1}^n \left [\tau\{\hat{F}_{(\data_i,\epsilon)}(\alpha)\}- \tau\{\hat{F}(\alpha)\}\right ]/{\epsilon} = 0 \label{eq:sum-gateaux}
\end{align}
with a small $\epsilon > 0$. The deductive estimator for $\tau = P(T > t)$ is $\hat\tau = \tau\{\hat{F}(\hat\alpha)\}$. The standard error of $\hat{\tau}$ can be estimated by $n^{-1}\big[ \sum_i \gateaux\{\data_i,$ $\hat{F}(\hat{\alpha}),\epsilon \}^2 \big]^{1/2}$.

Although formally the estimation procedure looks similar to that in \citet{frangakis2015deductive}, the underlying implementation is completely different. In the two-phase design considered in \citet{frangakis2015deductive}, they were able to derive analytically the estimand evaluated at a mixture distribution, $\tau\{\hat{F}_{(\data_i,\epsilon)}(\alpha)\}$, for any estimand $\tau$. For the double-sampling design we consider here, however, deriving $\tau\{\hat{F}_{(\data_i,\epsilon)}(\alpha)\}$ is complicated and error-prone because of the presence of survival outcomes. Our proposed discretized support separates the computation of the mixture distribution $\hat{F}_{(\data_i,\epsilon)}(\alpha)$ and the computation of the estimand for an arbitrary distribution $\tau(G)$, which makes the deductive estimation feasible for double-sampling designs.

The estimator $\hat\tau$ has the following robustness property. Suppose we decompose $F(\alpha) : =(R,H(\alpha))$, where $H$ isolates the components of the distribution that are modeled through $\alpha$ in step 3, and $R$ is the remaining part of $F$. Generally, the expected EIF, $\bar{\phi}_{(R_0,H_0)}(R,H):=E_{(R_0,H_0)} \{\phi(\data_i,R, H)\}$, is zero at the truth $(R_0,H_0)$ but possibly also at other values of $(R,H)$ (e.g., double-robustness). Under regularity conditions usually needed for consistency with estimating equations, the above estimator when taking a working model $R_w$ should be consistent if the model  $\{H(\alpha)\}$ includes a distribution $H^*$ that satisfies condition \eqref{condition} of Appendix \ref{appen:robust}
\begin{equation}
 \bar{\phi}_{(R_0,H_0)}(R_w,H^*)=0 \mbox{ and }\tau(R_w,H^*)= \tau(R_0,H_0),\nonumber
\end{equation}
that is, the distribution $H^*$ zeros out the limit EIF and gives the correct value of the estimand. 

\subsection{Working distributions $\hat{F}(\alpha)$ on the discretized support $\Omega$}
\label{subsec:F-alpha}

Here we give the form of the working distributions $\hat{F}(\alpha)$ used in Section \ref{subsec:algorithm}. We first describe how a working distribution $\hat{F}$ on $\Omega$ is estimated based on the observed data, then describe how to extend it to $\hat{F}(\alpha)$, a class of working distributions parameterized by a 1-dimensional $\alpha$.

In the following, let $o = (\robs, z, w, s, x, \delta)$ denote an arbitrary point in $\Omega$, let $P_{\hat{F}}(o)$ denote the probability of $o$ under the distribution $\hat{F}$, and let $\hat{P}$ denote a discrete probability measure that is estimated from the observed distribution, which we will describe after \eqref{eq:calculate-hatF}. The working distribution $\hat{F}$ is given by
\begin{align}
    P_{\hat{F}}(\robs, z, w, s, x, \delta) = & \hat{P}(\Robs = \robs) \hat{P}(Z=z, W=w \mid \Robs = \robs) \nonumber \\
        & \times \hat{P}(S=s \mid \Robs = \robs, Z=z, W=w) \nonumber \\
        & \times \hat{P}(X=x, \Delta=\delta \mid \Robs = \robs, Z=z, W=w, S=s). \label{eq:calculate-hatF}
\end{align}
For notation simplicity, we will omit the lowercase letters inside $\hat{P}(\cdot)$ for the rest of the section if no confusion is caused. For the terms on the right hand side of \eqref{eq:calculate-hatF}, $\hat{P}(\Robs)$ and $\hat{P}(Z, W \mid \Robs)$ are estimated by their empirical distributions. The selection probability for double-sampling, $\hat{P}(S \mid \Robs = 0, Z, W)$, is estimated by logistic regression. The probability on the survival information, $\hat{P}(X, \Delta \mid \Robs, Z, W, S)$, is estimated as follows.

By construction of $\Omega$, we have $(x,\delta) = (\na,\na)$ if and only if $\robs = 0, s = 0$, so $\hat{P}(X =\na, \Delta =\na \mid \Robs = 0, Z, W, S = 0) = 1$. For $\hat{P}(X, \Delta \mid \Robs, Z, W, S)$ with $(\Robs = 0,S=1)$ or $\Robs = 1$, we consider two different approaches: (i) an estimate based on Cox proportional hazards models, and (ii) an estimate based on log-normal regressions. For (i), Cox proportional hazards models \citep{cox1972regression} are fitted under the working independence assumption $T \ci C \mid \Robs, Z, W, S$ to obtain
\begin{align}
    \hat{P}(T = x \mid \Robs = 0, Z, W, S=1) \mbox{ and } \hat{P}(C = x \mid \Robs = 0, Z, W, S=1) &\mbox{ for } x\in \{ x_1,\ldots,x_{m_1} \}, \nonumber \\
    \hat{P}(T = x \mid \Robs = 1, Z, W, S=0) \mbox{ and } \hat{P}(C = x \mid \Robs = 1, Z, W, S=0) &\mbox{ for } x\in \{ x_{m+1},\ldots,x_n \}. \label{eq:prob-T-and-C}
\end{align}
For (ii), log-normal regressions are fitted under the working independence assumption $T \ci C \mid \Robs, Z, W, S$ to obtain probability density functions for $T \mid \Robs = 0, Z, W, S=1$, $C \mid \Robs = 0, Z, W, S=1$, $T \mid \Robs = 1, Z, W, S=0$, and $C \mid \Robs = 1, Z, W, S=0$. The probability density functions are then converted to \eqref{eq:prob-T-and-C}. For either approach, the fitted probability on $(X,\Delta)$ is then computed as
\begin{align}
    \hat{P}(X = x, \Delta = \delta \mid \Robs, Z, W, S) & = \big\{ \hat{P}(T=x \mid \Robs, Z, W, S) \hat{P}(C>x \mid \Robs, Z, W, S) \big\}^{\delta} \nonumber \\
    & \times \big\{ \hat{P}(T>x \mid \Robs, Z, W, S) \hat{P}(C=x \mid \Robs, Z, W, S) \big\}^{1 - \delta}, \label{eq:compute-prob-xdelta-from-tc}
\end{align}
and it is further normalized so that 
\begin{align}
    & \sum_{i=1}^{m_1} \hat{P}(X = x_i, \Delta = \delta_i \mid \Robs = 0, Z=z, W=w, S=1) = 1 \mbox{ if } \robs = 0, \nonumber \\
    & \sum_{i=m+1}^{n} \hat{P}(X = x_i, \Delta = \delta_i \mid \Robs = 1, Z=z, W=w) = 1 \mbox{ if } \robs = 1. \label{eq:normalize-prob-xdelta}
\end{align}
We note that in both approaches, the working models for $\Robs = 0$ and $\Robs = 1$ are variationally independent; i.e., no parameter is shared across working models. A more detailed exposition of how each term in \eqref{eq:calculate-hatF} is computed is given in Appendix \ref{appen:working-distributions}.

We then extend the working distribution $\hat{F}$ to $F(\alpha)$, a class of working distributions, by adding a 1-dimensional parameter $\alpha$ to $\hat{P}(X, \Delta \mid \Robs, Z, W, S)$ while leaving the other factors of $P_{\hat{F}}$ in \eqref{eq:calculate-hatF} unchanged, as follows. If $\hat{P}(X, \Delta \mid \Robs, Z, W, S)$ is obtained from Cox models (i.e., approach (i) above), the extended probability parameterized by $\alpha$ is:
\begin{align}
    & \hat{P}(X = x, \Delta = \delta \mid \Robs = \robs, Z, W, S; \alpha) \nonumber \\
    & \qquad \qquad = \max[0, \hat{P}(X = x, \Delta = \delta \mid \Robs = \robs, Z, W, S) \{ 1+ \alpha x / c_{\max}^{(\robs)}\}],  \label{eq:extended-working-model}
\end{align}
where $c_{\max}^{(0)} = \max\{ c_i: 1 \leq i \leq m \}$ and $c_{\max}^{(1)} = \max\{ c_i: m+1 \leq i \leq n \}$. For $\alpha>0$, this extension increases the probability of $(x,\delta)$ pairs with larger $x$ values; for $\alpha<0$, this extension increases the probability of $(x,\delta)$ pairs with smaller $x$ values; when $\alpha = 0$, this extension has no effect. The probability calculated by \eqref{eq:extended-working-model} is then normalized to ensure \eqref{eq:normalize-prob-xdelta}.

If $\hat{P}(X, \Delta \mid \Robs, Z, W, S)$ is obtained from log-normal regressions (i.e., approach (ii) above), suppose the fitted regression has mean (after log transformation)
\begin{align}
    \hat{E}\{\log(T) \mid \Robs = 0, Z, W, S=1\} &= \hat\beta_0 + Z\hat\beta_1 + W\hat\beta_2, \nonumber \\
    \hat{E}\{\log(T) \mid \Robs = 1, Z, W\} &= \hat\gamma_0 + Z\hat\gamma_1 + W\hat\gamma_2. \nonumber
\end{align}
The free parameter $\alpha$ is added to the intercepts $\hat\beta_0$ and $\hat\gamma_0$, and the extended probability $\hat{P}(X, \Delta \mid \Robs, Z, W, S; \alpha)$ is then computed using \eqref{eq:compute-prob-xdelta-from-tc}. For $\alpha>0$, this extension increases the probability for larger $T$ values; for $\alpha<0$, this extension increases the probability for smaller $T$ values; when $\alpha = 0$, this extension has no effect. Note that the log-normal regression fits of $C$ is unchanged.

\subsection{Remarks on the choice of $\Omega$ and $\hat{F}(\alpha)$}

We comment on our particular choice of the discretized support $\Omega$ and the working distributions $\hat{F}(\alpha)$, and discuss other choices that we have considered.

A necessary condition for $\Omega$ is to include all the observed data points to make the perturbation (i.e., $(1-\epsilon)\hat{F}(\alpha) + \epsilon \cL_{O_i}$) easy to compute. This is not sufficient, though. When we attempted to set $\Omega = \{O_1, \ldots, O_n\}$, even though the perturbation is straightforward, the support turned out to be not rich enough to allow the computation of the distribution components in \eqref{eq:distr1} and \eqref{eq:distr2} (especially the distributions involving $(X,\Delta)$). Therefore, we define $\Omega$ to be the Cartesian product of unique $(z,w)$ pairs and unique $(s,x,\delta)$ triples from the observed data, which allows computation of both the perturbation and the estimand. Note that instead of constructing an overall Cartesian product space, the discretized support is defined within $\Robs = 1$ and $\Robs = 0$, because $\pr(\Robs = 1)$ can be reliably estimated by the empirical frequency and we want the working distribution on $\Omega$ to accommodate this. We did not include $c_i$ in constructing $\Omega$, because as shown in \eqref{eq:estimand1} its distribution is not needed to identify the parameter of interest.

The choice of an appropriate $\Omega$ may depend on the dimension of $(Z,W)$. Suppose, for example, that $(Z,W)$ can only take a few categorical values, then $\Omega' = \{O_1, \ldots, O_n\}$ may be rich enough to allow computation of the distribution components in \eqref{eq:distr1} and \eqref{eq:distr2} when the sample size is relatively large compared to the number of $(Z,W)$-categories. However, when $(Z,W)$ is multi-dimensional or has continuous variables, a richer discretized support such as the Cartesian product used in Section \ref{subsec:algorithm} is needed. It is an open question regarding the exact requirements for the discretized support $\Omega$, and how such requirements relate to the dimensionality of the variables involved and to the sample size of the data.

We described in Section \ref{subsec:F-alpha} two approaches to fit the working distribution $\hat{F}$ on $\Omega$, one based on Cox proportional hazards models and the other based on log-normal regressions. Extending $\hat{F}$ to $\hat{F}(\alpha)$ is different for the two approaches: for Cox model-based $\hat{F}$, the free parameter $\alpha$ is added onto the probability distribution of $(X,\Delta)$ directly; for log-normal regression-based $\hat{F}$, the free parameter $\alpha$ is added as an intercept to the regression fits for $T$. A necessary condition for a valid extension $\hat{F}(\alpha)$ is that \eqref{eq:sum-gateaux} explores values around 0 as $\alpha$ varies. Other attempts, such as adding the free parameter $\alpha$ to a regression coefficient in the Cox model fits of $T$ and $C$ for Cox model-based $\hat{F}$, have failed due to the fact that \eqref{eq:sum-gateaux} does not equal zero for any $\alpha$ value.

We note that the working independence assumption $T \ci C \mid \Robs, Z, W, S$ and other modeling assumptions used in calculating $\hat{P}(X, \Delta \mid \Robs, Z, W, S)$ (proportional hazards assumption for Cox model-based $\hat{F}$, parametric form for log-normal regression-based $\hat{F}$) may not and is not expected to hold, and they are just a convenient way to construct $\hat{P}(X, \Delta \mid \Robs, Z, W, S)$. In fact, \citet{frangakis2001} showed in a setting without covariates that $T \ci C \mid \Robs$ can fail to hold. In the following simulation studies, we see that the proposed estimators are robust to violation of those working assumptions.

\section{Simulation}
\label{sec:simulation}

We evaluate the performance of the proposed deductive estimators and compare it with other existing approaches through simulation studies. In Section \ref{subsec:simulation-gm} we describe the generative models used in the simulations. In Section \ref{subsec:simulation-consistency} the proposed estimators are compared with three other estimators in terms of consistency and coverage probability of confidence intervals. In Section \ref{subsec:simulation-incorrect-S} we evaluate the performance of the proposed estimators when an incorrect model for the double-sampling selection is used. In Section \ref{subsec:simulation-alpha} we assess the impact of extending the working model $\hat{F}$ to $\hat{F}(\alpha)$ on the performance of the proposed estimators.

The simulation results show that the proposed estimators are consistent under possibly incorrect working models for both $(X,\Delta)$ and for $S$. The confidence interval can be anti-conservative for small sample sizes, and it becomes close to the nominal level coverage as sample size gets larger. The simulation also shows that the model extension step (from $\hat{F}$ to $\hat{F}(\alpha)$) contributes to the robustness of the estimators.

We briefly summarize the results of some additional simulation studies, the details of which are not listed in the paper. The proposed estimators are not sensitive to the choice of $\epsilon$ as long as $\epsilon$ is small enough (recall that $\epsilon$ is used in calculating the numerical Gateaux derivative in \eqref{eq:sum-gateaux}): we varied $\epsilon$ from $1\times 10^{-4}$ to $1\times 10^{-8}$ given the sample sizes used in the simulation study ($n=50$, $100$, $200$, $500$), and the performance of the proposed estimators stays effectively the same.  In addition to the continuous generative model described below, we used a discrete generative model (where $Z,T,C,L$ are all discrete and each has around $10$ levels), and the proposed estimators are consistent with nominal level confidence interval coverage.

\subsection{Generative models}
\label{subsec:simulation-gm}

We consider two generative models. In the first generative model (GM-1), the baseline covariate, $Z$, is generated from $\mbox{Uniform}[-2, 2]$. The true dropout status, $R$, is generated as Bernoulli random variables with success probability $(Z+3)/6$. The survival time, $T$, is generated as $5$ times a Weilbull-distributed random variable with scale parameter $\exp(Z)$ and shape parameter $5$. The administrative censoring time, $C$, is generated from $\mbox{Uniform}[0.5, 2]$, which mimics a continuous accrual process with a constant accrual rate. The dropout time, $L$, is generated from $\mbox{Uniform}[T/3, T]$ for those with $R=0$, and is $\na$ for those with $R=1$. This is the only longitudinal measurement; i.e., $W = L$.  The observed dropout status, $\Robs$, equals 1 for those with $R=1$, and is defined as $\Robs = \indic\{\min(T,C) < L\}$ for those with $R=0$, where $\indic(\cdot)$ is the indicator function. $L$ is then set to $\na$ for those with $(R=0, \Robs = 1)$. The selection for double-sampling, $S$, is generated as a Bernoulli random variable with success probability $\mbox{expit}\{(L+Z+1)/2\}$ for those with $\Robs = 0$, where $\mbox{expit}(y) = \{1+\exp(-y)\}^{-1}$. The survival data, $(X,\Delta)$, is defined by $X = \min(T,C)$ and $\Delta = \indic(T \leq C)$ for those with $\Robs = 1$ or $(\Robs=0,S=1)$, and is $(\na,\na)$ otherwise. The parameter of interest is $\tau = P(T > 0.7) = 77.1\%$. 

In the second generative model (GM-2), $Z$ and $R$ are generated the same way as in GM-1. The survival time $T$ and the administrative censoring time $C$ are generated independently from a log-normal distribution $\mbox{LN(Z, 0.25)}$, where $A \sim \mbox{LN}(\mu, \sigma^2)$ means that $\log(A)$ follows a normal distribution with mean $\mu$ and variance $\sigma^2$. The dropout time $L \sim \mbox{Uniform}[T/3, T]$ for those with $R=0$ and is $\na$ for those with $R=1$. $\Robs, X, \Delta$ are calculated the same way as in GM-1, and $S$ is generated as Bernoulli with success probability $\mbox{expit}\{(L-Z+1)/2\}$ for those with $\Robs = 0$.  The parameter of interest is $\tau = P(T > 0.7) = 58.8\%$. 

\begin{table}[htbp]
\begin{center}
\begin{tabular}{lccc}
\toprule
                                                             & & GM-1         & GM-2         \\ 
\midrule
$\tau = P(T > 0.7)$                                          & & 77.1\%         & 58.8\%         \\
$P(\Robs = 0)$                                               & & 0.43         & 0.36         \\
$P(S=1 \mid \Robs = 0)$                                      & & 0.65         & 0.73         \\
10-th and 90-th percentiles of $P(S=1 \mid \Robs = 0, Z, L)$ & & (0.50, 0.80) & (0.65, 0.81) \\
$P(\Delta = 1 \mid \Delta \neq \na)$                         & & 0.70         & 0.48         \\
10-th and 90-th percentiles of $X \mid X \neq \na$           & & (0.53, 1.18) & (0.15, 3.59)   \\
$\mbox{corr}(T, C \mid Z, \Robs)$                            & & 0.02         & 0.24         \\
$\mbox{corr}(T, C \mid Z, L, \Robs = 0)$                     & & 0.00         & 0.12         \\ \bottomrule
\end{tabular}    
\end{center}
\caption{Descriptive statistics of the two generative models. $\mbox{corr}$ stands for Pearson's partial correlation. All quantities are calculated from a simulated data set with 1,000,000 sample size.}
\label{tab:GM}
\end{table}

Table \ref{tab:GM} lists the descriptive statistics of the two generative models, which are calculated from a simulated data set with 1,000,000 sample size for each GM. For GM-1 (GM-2), $43\%$ ($36\%$) individuals drop out during the study, and among them $65\%$ ($73\%$) are double-sampled. For the patients with $\Robs = 0$, the 10-th and the 90-th percentiles of the probability of being double-sampled are 0.50 and 0.80 for GM-1 (0.65 and 0.81 for GM-2). Among the patients with observed $(X,\Delta)$, $70\%$ are dead on observation ($\Delta = 1$) for GM-1 ($48\%$ for GM-2), and the 10-th and the 90-th percentiles of $X$ are 0.53 and 1.18 for GM-1 (0.15 and 3.59 for GM-2). The Pearson's partial correlation is $\mbox{corr}(T, C \mid Z, \Robs) = 0.02$ for GM-1 (0.24 for GM-2), and $\mbox{corr}(T, C \mid Z, L, \Robs = 0) = 0.00$ for GM-1 (0.12 for GM-2); these are computed using \textsf{R} package \textsf{ppcor} \citep{Rppcor}. This implies that although Assumption \ref{assumpt:independent-censoring} holds for both GMs, independent censoring conditional on $Z,\Robs$ or $Z,L,\Robs$ does not hold for GM-2 and may be slightly violated for GM-1. In addition, in GM-1 the proportional hazards assumption on $T\mid Z$ holds, and in GM-2 the proportional hazards assumption on $T\mid Z$ does not hold \citep{bender2005coxph}.

\subsection{Simulation on consistency}
\label{subsec:simulation-consistency}

We evaluate the performance of the proposed estimators in terms of consistency and confidence interval coverage and compare them with other existing methods. We consider the following five estimators:
\begin{itemize}
    \item DE.Cox: the deductive estimator proposed in Section \ref{sec:method} with Cox model as working model for the distribution of $(X,\Delta)$. $(Z,L)$ are included in both the Cox model fits and the model for double-sampling selection $S$. $\epsilon$ is set to be $1\times 10^{-4}$.
    \item DE.LN: the deductive estimator proposed in Section \ref{sec:method} with log-normal regression as working model for the distribution of $(X,\Delta)$. $(Z,L)$ are included in both the log-normal regression and the model for double-sampling selection $S$.  $\epsilon$ is set to be $1\times 10^{-4}$.
    \item PAR: the estimator in \citet{an2014statmed} that is based on parametric assumptions. In this particular implementation we used, $T \mid R=1, Z$ and $T \mid R=0, Z, L$ are assumed to follow log-normal distributions. The \textsf{R} code to compute the estimator and the standard error is provided by Dr. Ming-Wen An.
    \item KM.S: a stratified Kaplan-Meier estimator with stratification variable $\Robs$. In this estimator, a Kaplan-Meier estimator \citep{kaplan1958nonparametric} is computed separately for the patients with $(\Robs=0,S=1)$ and for the patients with $\Robs=1$, and a weighted average is taken using weights $\hat{P}(\Robs)$. The bias-corrected and accelerated (BCa) bootstrap 95\% confidence interval is obtained using \textsf{R} package \textsf{boot} \citep{Rboot}, with 1000 bootstrap replicates.
    \item KM.C: a Kaplan-Meier estimator using only complete cases (i.e., data with $\Robs = 1$ or $(\Robs = 0, S = 1)$) without any weighting.
\end{itemize}

\begin{table}[htbp]

\begin{center}
\resizebox{\linewidth}{!}{
\begin{tabular}[t]{ccccccccccccccccc}
\toprule
\multicolumn{1}{c}{ } & \multicolumn{1}{c}{ } & \multicolumn{3}{c}{DE.Cox} & \multicolumn{3}{c}{DE.LN} & \multicolumn{3}{c}{PAR} & \multicolumn{3}{c}{KM.S} & \multicolumn{3}{c}{KM.C} \\
\cmidrule(l{2pt}r{2pt}){3-5} \cmidrule(l{2pt}r{2pt}){6-8} \cmidrule(l{2pt}r{2pt}){9-11} \cmidrule(l{2pt}r{2pt}){12-14} \cmidrule(l{2pt}r{2pt}){15-17}
GM & $n$ & Bias & CP & SD & Bias & CP & SD & Bias & CP & SD & Bias & CP & SD & Bias & CP & SD\\
\midrule
 & 50 & 0.6 & 92.1 & 6.5 & 0.0 & 91.6 & 6.7 & -1.5 & 89.9 & 6.0 & -0.8 & 95.6 & 6.9 & -1.4 & 97.7 & 6.8\\

 & 100 & -0.1 & 94.2 & 4.6 & -0.2 & 93.7 & 4.6 & -1.9 & 89.4 & 4.2 & -1.1 & 95.1 & 4.9 & -1.7 & 98.6 & 4.9\\

 & 200 & 0.2 & 93.2 & 3.3 & 0.1 & 92.7 & 3.3 & -1.9 & 87.4 & 3.0 & -0.8 & 93.9 & 3.4 & -1.4 & 98.8 & 3.4\\

\multirow{-4}{*}{\centering\arraybackslash 1} & 500 & 0.0 & 93.9 & 2.0 & 0.0 & 93.6 & 2.0 & -2.1 & 74.8 & 1.9 & -0.9 & 93.0 & 2.2 & -1.5 & 98.1 & 2.2\\
\cmidrule{1-17}
 & 50 & 2.3 & 85.6 & 9.2 & 0.4 & 88.7 & 11.6 & 0.2 & 92.8 & 7.2 & 12.7 & 67.4 & 7.8 & 15.0 & 62.6 & 7.5\\

 & 100 & 0.3 & 89.8 & 6.2 & 0.7 & 91.1 & 7.1 & -0.2 & 94.2 & 4.7 & 12.9 & 36.9 & 5.3 & 15.2 & 37.6 & 5.0\\

 & 200 & -0.2 & 92.1 & 4.1 & 0.7 & 93.4 & 4.5 & -0.2 & 94.4 & 3.3 & 12.9 & 8.7 & 3.7 & 15.2 & 8.8 & 3.5\\

\multirow{-4}{*}{\centering\arraybackslash 2} & 500 & -0.4 & 92.8 & 2.5 & 0.4 & 93.8 & 2.7 & -0.4 & 94.7 & 2.0 & 12.6 & 0.1 & 2.3 & 14.9 & 0.1 & 2.3\\
\bottomrule
\end{tabular}}
\end{center}

\vspace{-0.8em}
\footnotesize{* Note: Bias and SD in the table have been multiplied by 100, so the Bias and SD are on the scale of percentage survival probability.}

\caption{Simulation result for estimators DE.Cox, DE.LN, PAR, KM.S aand KM.C. GM: generative model; $n$: sample size; CP: coverage probability (in \%) of 95\% confidence interval based on normal approximation; SD: standard deviation. All results are based on 1000 replicates. }
\label{tab:simulation-consistency}
\end{table}

The simulation result is shown in Table \ref{tab:simulation-consistency}. Simulation is conducted for sample size $n=50$, $100$, $200$, $500$, each with $1000$ replicates. Both DE.Cox and DE.LN are consistent under both GM-1 and GM-2; their coverage probability of $95\%$ confidence interval based on normal approximation is anti-conservative for small sample sizes such as $n=50$ (with actual coverage probability being around $92\%$), but it becomes close to the nominal level for $n=500$. PAR is consistent under GM-2 because $T \mid R, Z$ follows log-normal distribution, and it is more efficient (i.e., has smaller standard deviation) than the other estimators because of the parametric assumption; it is inconsistent under GM-1 because the log-normal parametric assumption is violated there. KM.S is possibly inconsistent under GM-1, as the coverage probability starts to decrease as $n$ gets larger; however, this possible inconsistency is small because $\mbox{corr}(T, C \mid Z, \Robs) = 0.02$, which means that the assumption $T \ci C \mid Z, \Robs$ required for its consistency is only slightly violated. KM.S is severely inconsistent under GM-2, where $\mbox{corr}(T, C \mid Z, \Robs) = 0.24$. KM.C is severely inconsistent under GM-2; under GM-1, it is possibly inconsistent because the bias does not decrease as $n$ grows, but its confidence interval is conservative and has larger than nominal coverage probability.

The simulation result shows the potential robustness of DE.Cox and DE.LN. Under GM-2, the working independence assumption $T \ci C \mid Z,L,\Robs$ and the proportional hazards assumption for $T \mid Z,L,\Robs$ are both likely violated, yet DE.Cox is consistent with close to nominal confidence interval coverage especially when $n$ is large. Under GM-1, the log-normal distributional assumption $T \mid Z,L,\Robs$ is likely violated, yet DE.LN is consistent with close to nominal confidence interval coverage especially when $n$ is large.

\subsection{Simulation with incorrect double-sampling model for $S$}
\label{subsec:simulation-incorrect-S}

We evaluate the performance of DE.Cox and DE.LN when an incorrect model for the double-sampling selection, $P(S=1 \mid \Robs = 0, Z, L)$, is used. In particular, let DE.Cox.WrongS and DE.LN.WrongS denote the estimators DE.Cox and DE.LN, respectively, when the model $\hat{P}(S=1 \mid \Robs = 0, Z, L)$ only includes an intercept (i.e., regressors $Z,L$ are omitted). The simulation result is shown in Table \ref{tab:simulation-incorrect-S}. For both estimators, the performances under correct model for $S$ and under incorrect model for $S$ are comparable under both GMs, indicating that the proposed estimators are likely robust to incorrect models of the double-sampling selection.

\begin{table}[htbp]

\begin{center}
\resizebox{\linewidth}{!}{
\begin{tabular}[t]{cccccccccccccc}
\toprule
\multicolumn{1}{c}{ } & \multicolumn{1}{c}{ } & \multicolumn{3}{c}{DE.Cox} & \multicolumn{3}{c}{DE.Cox.WrongS} & \multicolumn{3}{c}{DE.LN} & \multicolumn{3}{c}{DE.LN.WrongS} \\
\cmidrule(l{2pt}r{2pt}){3-5} \cmidrule(l{2pt}r{2pt}){6-8} \cmidrule(l{2pt}r{2pt}){9-11} \cmidrule(l{2pt}r{2pt}){12-14}
GM & $n$ & Bias & CP & SD & Bias & CP & SD & Bias & CP & SD & Bias & CP & SD\\
\midrule
 & 50 & 0.6 & 92.1 & 6.5 & 0.4 & 93.3 & 6.5 & 0.0 & 91.6 & 6.7 & 0.1 & 92.3 & 6.7\\

 & 100 & -0.1 & 94.2 & 4.6 & 0.0 & 94.8 & 4.5 & -0.2 & 93.7 & 4.6 & -0.1 & 94.0 & 4.5\\

 & 200 & 0.2 & 93.2 & 3.3 & 0.3 & 94.2 & 3.3 & 0.1 & 92.7 & 3.3 & 0.3 & 93.1 & 3.3\\

\multirow{-4}{*}{\centering\arraybackslash 1} & 500 & 0.0 & 93.9 & 2.0 & 0.2 & 94.5 & 2.0 & 0.0 & 93.6 & 2.0 & 0.1 & 93.9 & 2.0\\
\cmidrule{1-14}
 & 50 & 2.3 & 85.6 & 9.2 & 2.3 & 85.9 & 9.3 & 0.4 & 88.7 & 11.6 & 0.4 & 88.8 & 11.6\\

 & 100 & 0.3 & 89.8 & 6.2 & 0.3 & 89.7 & 6.2 & 0.7 & 91.1 & 7.1 & 0.7 & 90.9 & 7.1\\

 & 200 & -0.2 & 92.1 & 4.1 & -0.2 & 92.2 & 4.1 & 0.7 & 93.4 & 4.5 & 0.7 & 93.1 & 4.5\\

\multirow{-4}{*}{\centering\arraybackslash 2} & 500 & -0.4 & 92.8 & 2.5 & -0.4 & 92.8 & 2.5 & 0.4 & 93.8 & 2.7 & 0.4 & 93.8 & 2.7\\
\bottomrule
\end{tabular}}
\end{center}

\vspace{-0.8em} \footnotesize{* Note: Bias and SD in the table have been multiplied by 100, so the Bias and SD are on the scale of percentage survival probability.}

\caption{Simulation result for estimators DE.Cox and DE.LN, and when they are using incorrect model for $P(S=1\mid \Robs = 0, Z, W)$ (columns with WrongS). GM: generative model; $n$: sample size; CP: coverage probability (in \%) of 95\% confidence interval based on normal approximation; SD: standard deviation. All results are based on 1000 replicates. The columns for DE.Cox and DE.LN are the same as those in Table \ref{tab:simulation-consistency}, and are included here for comparison.}
\label{tab:simulation-incorrect-S}
\end{table}

\subsection{Simulation on the impact of $\alpha$}
\label{subsec:simulation-alpha}

We evaluate the impact of the free parameter $\alpha$ on the performance of the proposed estimators. In particular, we consider two additional estimators: DE.Cox($\alpha=0$) and DE.LN($\alpha=0$). DE.Cox($\alpha=0$) uses the same working models as DE.Cox, but it did not solve \eqref{eq:sum-gateaux} for $\hat{\alpha}$ and output as estimator $\tau\{\hat{F}(\hat\alpha)\}$; instead, it sets $\alpha = 0$ and outputs as estimator $\tau\{\hat{F}(0)\}$. DE.LN($\alpha=0$) uses the same working models as DE.LN, but it sets $\alpha = 0$ and outputs as estimator $\tau\{\hat{F}(0)\}$. Recall that for DE.Cox and DE.LN, $\alpha = 0$ means that no extension is added to the fitted working distribution $\hat{F}$.

\begin{table}[htbp]

\begin{center}
\resizebox{\linewidth}{!}{
\begin{tabular}[t]{cccccccccccccc}
\toprule
\multicolumn{1}{c}{ } & \multicolumn{1}{c}{ } & \multicolumn{3}{c}{DE.Cox} & \multicolumn{3}{c}{DE.Cox($\alpha=0$)} & \multicolumn{3}{c}{DE.LN} & \multicolumn{3}{c}{DE.LN($\alpha=0$)} \\
\cmidrule(l{2pt}r{2pt}){3-5} \cmidrule(l{2pt}r{2pt}){6-8} \cmidrule(l{2pt}r{2pt}){9-11} \cmidrule(l{2pt}r{2pt}){12-14}
GM & $n$ & Bias & CP & SD & Bias & CP & SD & Bias & CP & SD & Bias & CP & SD\\
\midrule
 & 50 & 0.6 & 92.1 & 6.5 & 1.1 & 92.8 & 6.3 & 0.0 & 91.6 & 6.7 & -2.9 & 83.7 & 8.3\\

 & 100 & -0.1 & 94.2 & 4.6 & 0.3 & 94.2 & 4.5 & -0.2 & 93.7 & 4.6 & -3.3 & 82.1 & 5.9\\

 & 200 & 0.2 & 93.2 & 3.3 & 0.3 & 93.4 & 3.3 & 0.1 & 92.7 & 3.3 & -3.1 & 76.9 & 4.1\\

\multirow{-4}{*}{\centering\arraybackslash 1} & 500 & 0.0 & 93.9 & 2.0 & 0.0 & 94.3 & 2.0 & 0.0 & 93.6 & 2.0 & -3.5 & 55.7 & 2.6\\
\cmidrule{1-14}
 & 50 & 2.3 & 85.6 & 9.2 & 5.2 & 82.7 & 9.5 & 0.4 & 88.7 & 11.6 & 3.5 & 85.8 & 10.4\\

 & 100 & 0.3 & 89.8 & 6.2 & 1.5 & 89.4 & 6.1 & 0.7 & 91.1 & 7.1 & 2.6 & 89.8 & 6.9\\

 & 200 & -0.2 & 92.1 & 4.1 & 0.3 & 93.5 & 4.0 & 0.7 & 93.4 & 4.5 & 2.6 & 89.9 & 4.7\\

\multirow{-4}{*}{\centering\arraybackslash 2} & 500 & -0.4 & 92.8 & 2.5 & -0.5 & 92.9 & 2.5 & 0.4 & 93.8 & 2.7 & 2.1 & 84.4 & 2.9\\
\bottomrule
\end{tabular}}
\end{center}

\vspace{-0.8em} \footnotesize{* Note: Bias and SD in the table have been multiplied by 100, so the Bias and SD are on the scale of percentage survival probability.}

\caption{Simulation result for estimators DE.Cox, DE.Cox($\alpha=0$), DE.LN, and DE.LN($\alpha=0$). GM: generative model; $n$: sample size; CP: coverage probability (in \%) of 95\% confidence interval based on normal approximation; SD: standard deviation. All results are based on 1000 replicates. The columns for DE.Cox and DE.LN are the same as those in Table \ref{tab:simulation-consistency}, and are included here for comparison.}
\label{tab:simulation-alpha}
\end{table}

The simulation result is shown in Table \ref{tab:simulation-alpha}. DE.Cox and DE.Cox($\alpha=0$) have very similar performance under both GMs. On the other hand, DE.LN and DE.LN($\alpha=0$) have very different performance: DE.LN is consistent, whereas DE.LN($\alpha=0$) is severely inconsistent under both GMs. This shows the robustness brought by the $\hat\alpha$ that solves \eqref{eq:sum-gateaux}.

\section{Application to PEPFAR}
\label{sec:application}


We apply the proposed method to estimating the mortality rate using the data set from a double-sampling design component of the President's Emergency Plan for Aids Relief (PEPFAR), and HIV monitoring and treatment program in East Africa evaluating the antiretroviral treatment (ART) for HIV-infected people \citep{geng2015ART}. The data set consists of 1,773 HIV-infected adults from Morogoro, Tanzania, who started ART after entering the study. 673 patients (38\%) dropped out during the study. Among the dropouts, 91 (14\% of the dropouts) got double-sampled. We use baseline age and pre-treatment CD4 value as $\infoE$, and the loss to follow-up time $L$ and the CD4 value measured at the last visit before dropout as $\infoL$. The first column in Figure \ref{fig:design} gives the number of patients of each type; for the types that are not distinguishable from the observed data (such as (b1) and (b2)), a single number is reported.

\begin{figure}[htbp]
\centering
\begin{minipage}{.48\textwidth}
  \centering
  \includegraphics[width=\linewidth]{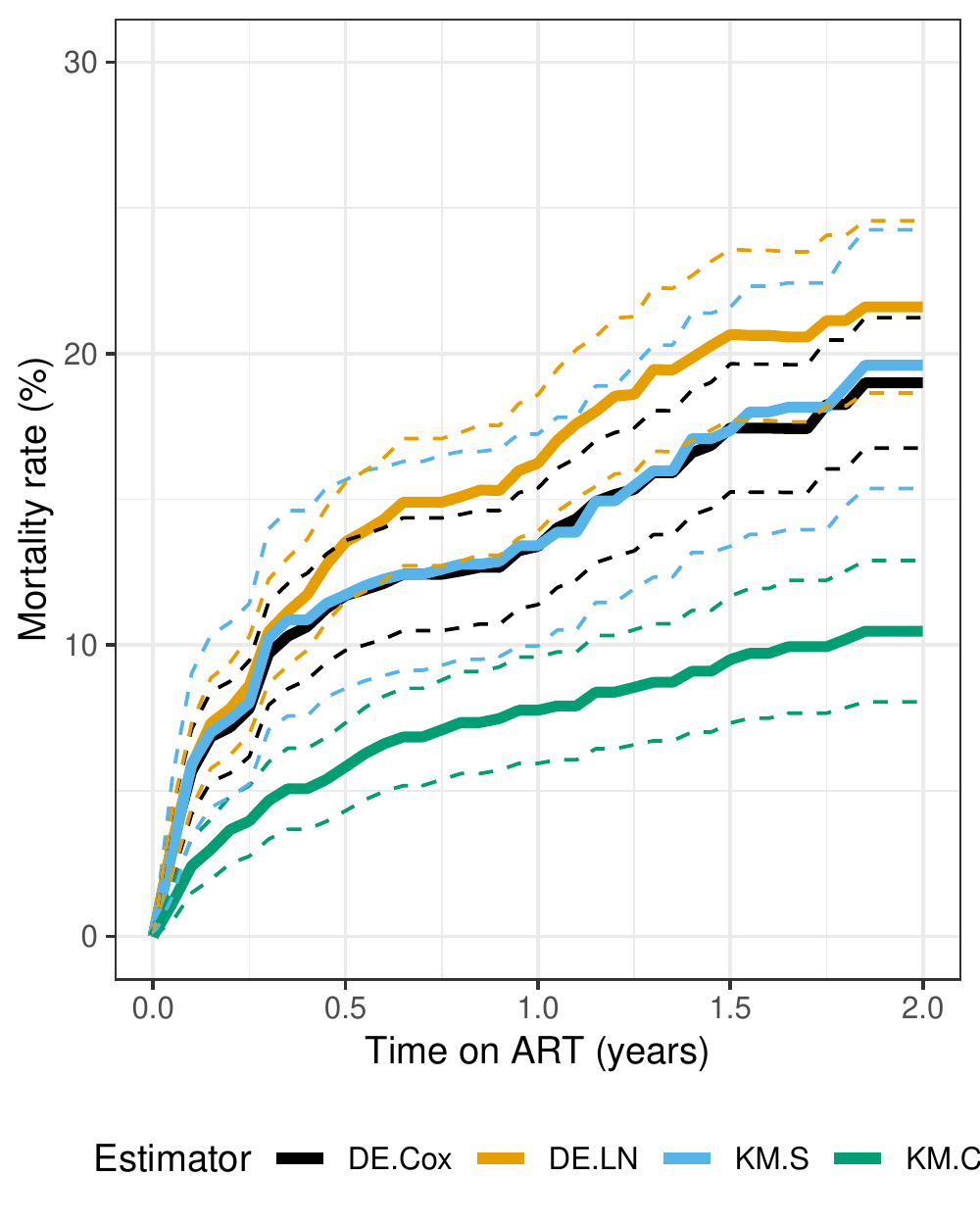}
  \caption{Estimated mortality rate (solid line) with 95\% pointwise confidence interval (dashed line) using all patients data from PEPFAR with four estimators: DE.Cox, DE.LN, KM.S, KM.C. \protect\phantom{These text are intended to make the two plots align with each other,}
  \protect\phantom{by making the number of lines of the caption equal.}
  \protect\phantom{Yet another line..............................}}
  \label{fig:PEPFAR-mortality-4est}
\end{minipage}
\hfill
\begin{minipage}{.48\textwidth}
  \centering
  \includegraphics[width=\linewidth]{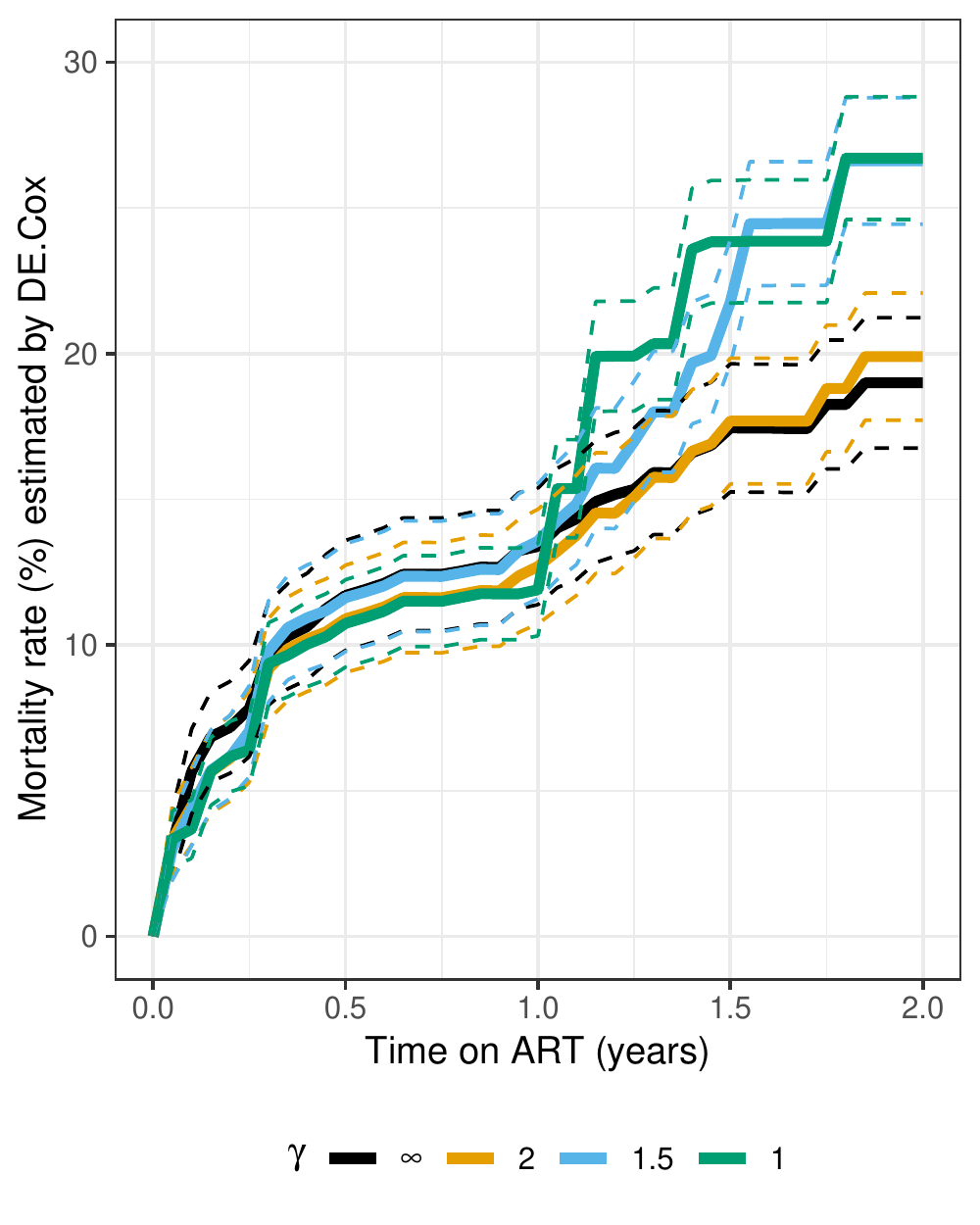}
  \caption{Estimated mortality rate (solid line) with 95\% pointwise confidence interval (dashed line) estimated by DE.Cox, using PEPFAR patients data with various $\gamma$ restrictions: all double-samples included ($\gamma = \infty$), past 2-year dropouts included ($\gamma = 2$), past 1.5-year dropouts included ($\gamma = 1.5$), past 1-year dropouts included ($\gamma = 1$).}
  \label{fig:PEPFAR-mortality-gamma}
\end{minipage}
\end{figure}

\bigskip
\noindent \textit{Estimated mortality rate.} We use the estimators proposed in Section \ref{sec:method} and two other estimators considered in the simulation study (Section \ref{subsec:simulation-consistency}) to estimate the mortality rate $P(T \leq t)$ for $0 \leq t \leq 2$ using the PEPFAR data. Figure \ref{fig:PEPFAR-mortality-4est} shows the result, with solid curves representing the estimates and dashed curves representing the 95\% pointwise confidence intervals. The deductive estimator using Cox proportional hazards model (DE.Cox, in black) and the stratified Kaplan-Meier estimator (KM.S, in blue) give very similar mortality rate estimates, although the confidence interval of KM.S is wider. The deductive estimator using log-normal regression (DE.LN, in yellow) gives slightly higher mortality rate estimates, with confidence interval width comparable to DE.Cox. The complete-case Kaplan-Meier estimator (KM.C, in green) gives much lower estimates than the other three; it is known to be biased for double-sampling designs \citep{frangakis2001}, and we include it for reference. The estimates and the confidence intervals for $t=0.5$, $1$, $1.5$, $2$ are listed in the first four rows (with $\gamma = \infty$) in Table \ref{tab:data-analysis-mortality-4est}.

\begin{table}[htbp]
\begin{center}
\resizebox{\linewidth}{!}{
\begin{tabular}[t]{cccccccccc}
\toprule
\multicolumn{1}{c}{ } & \multicolumn{1}{c}{ } & \multicolumn{2}{c}{DE.Cox} & \multicolumn{2}{c}{DE.LN} & \multicolumn{2}{c}{KM.S} & \multicolumn{2}{c}{KM.C} \\
\cmidrule(l{2pt}r{2pt}){3-4} \cmidrule(l{2pt}r{2pt}){5-6} \cmidrule(l{2pt}r{2pt}){7-8} \cmidrule(l{2pt}r{2pt}){9-10}
$\gamma$ & $t$ & mort & CI & mort & CI & mort & CI & mort & CI\\
\midrule
 & 0.5 & 11.7 & (9.8, 13.6) & 13.5 & (11.5, 15.6) & 11.7 & (8.5, 15.7) & 5.8 & (4.3, 7.3)\\

 & 1.0 & 13.4 & (11.4, 15.4) & 16.2 & (13.9, 18.6) & 13.4 & (10.0, 17.2) & 7.8 & (5.9, 9.6)\\

 & 1.5 & 17.5 & (15.3, 19.7) & 20.7 & (17.7, 23.6) & 17.4 & (13.4, 21.6) & 9.5 & (7.3, 11.7)\\

\multirow{-4}{*}{\centering\arraybackslash $\infty$} & 2.0 & 19.0 & (16.8, 21.2) & 21.6 & (18.7, 24.6) & 19.6 & (15.4, 24.2) & 10.5 & (8.1, 12.9)\\
\cmidrule{1-10}
 & 0.5 & 10.9 & (9.1, 12.7) & 13.4 & (11.4, 15.4) & 10.4 & (7.3, 14.7) & 5.3 & (3.9, 6.8)\\

 & 1.0 & 12.7 & (10.7, 14.7) & 16.4 & (14.0, 18.7) & 12.2 & (8.7, 16.2) & 7.3 & (5.5, 9.1)\\

 & 1.5 & 17.7 & (15.5, 19.8) & 21.3 & (18.3, 24.3) & 16.9 & (12.8, 21.8) & 9.1 & (6.9, 11.2)\\

\multirow{-4}{*}{\centering\arraybackslash 2} & 2.0 & 19.9 & (17.7, 22.1) & 22.3 & (19.3, 25.3) & 19.8 & (15.3, 25.5) & 10.1 & (7.7, 12.5)\\
\cmidrule{1-10}
 & 0.5 & 11.6 & (9.8, 13.5) & 15.3 & (13.2, 17.4) & 11.7 & (7.7, 16.3) & 5.3 & (3.9, 6.8)\\

 & 1.0 & 13.6 & (11.6, 15.5) & 18.8 & (16.3, 21.3) & 13.6 & (9.5, 18.3) & 7.3 & (5.5, 9.1)\\

 & 1.5 & 21.8 & (19.7, 23.9) & 25.0 & (21.9, 28.1) & 19.9 & (14.9, 25.6) & 9.2 & (7.0, 11.3)\\

\multirow{-4}{*}{\centering\arraybackslash 1.5} & 2.0 & 26.6 & (24.4, 28.8) & 27.5 & (24.4, 30.5) & 24.4 & (18.3, 31.0) & 10.2 & (7.7, 12.7)\\
\cmidrule{1-10}
 & 0.5 & 10.7 & (9.2, 12.2) & 16.6 & (14.6, 18.6) & 11.7 & (6.7, 17.8) & 4.8 & (3.4, 6.2)\\

 & 1.0 & 11.9 & (10.3, 13.5) & 19.6 & (17.3, 22.0) & 13.0 & (7.9, 19.1) & 6.7 & (4.9, 8.4)\\

 & 1.5 & 23.8 & (21.7, 25.9) & 23.9 & (21.1, 26.7) & 18.6 & (11.6, 26.8) & 7.8 & (5.8, 9.9)\\

\multirow{-4}{*}{\centering\arraybackslash 1} & 2.0 & 26.7 & (24.6, 28.8) & 25.7 & (22.9, 28.4) & 21.1 & (13.7, 29.6) & 8.4 & (6.2, 10.5)\\
\bottomrule
\end{tabular}}
\end{center}
\caption{Estimated $t$-year mortality rate (mort, in \%) and 95\% confidence intervals (CI) using estimators DE.Cox, DE.LN, KM.S, and KM.C. $\gamma = \infty$ means that the estimates are calculated from the original PEPFAR data. $\gamma = 2$, $1.5$, or $1$ means that the estimates are calculated from the pseudo data set where only the past 2-, 1.5-, or 1-year dropouts are possible for double-sampling.}
\label{tab:data-analysis-mortality-4est}
\end{table}

\bigskip
\noindent \textit{Impact of the double-sampling selection criteria on the estimated mortality rate.} In practice, to double-sample a dropout may require a considerable amount of effort and resource; therefore, it may be more feasible to double-sample relatively recent dropouts \citep{an2014statmed}. We evaluate the impact of such restrictions on double-sampling selection on the estimated mortality, by creating pseudo data sets based on the PEPFAR data, as follows. Suppose an investigator decides that only the \textit{recent} dropouts have positive probability to be double-sampled, where a dropout is \textit{recent} if his/her dropout time is within the past $\gamma$ years; in other words, if $C_i - L_i \leq \gamma$. A pseudo data set with regard to $\gamma$ is created by setting $S_i = 0$ and $(X_i, \Delta_i) = (\na, \na)$ for those in the PEPFAR data with $C_i - L_i > \gamma$. Figure \ref{fig:PEPFAR-mortality-gamma} gives the estimated morality rates (in solid curves) and the 95\% pointwise confidence intervals (in dashed curves) for the pseudo data sets with different $\gamma$, using the deductive estimator with Cox proportional hazards model (DE.Cox). Four $\gamma$ values are considered: $\gamma = \infty$ (no restrictions on double-sampling selection, which corresponds to 91 double-samples in the original PEPFAR data), and $\gamma = 2$, $1.5$, $1$ (only the past 2-, 1.5-, or 1-year dropouts are possible for double-sampling, which corresponds to 77, 62, or 33 double-samples in the pseudo data set). By definition, the curve for $\gamma = \infty$ (black curve in Figure \ref{fig:PEPFAR-mortality-gamma}) is the same as the black curve in Figure \ref{fig:PEPFAR-mortality-4est}. We see that for DE.Cox, the curve for $\gamma = 2$ (in yellow) is similar to that for $\gamma = \infty$ throughout $0\leq t \leq 2$; the curve for $\gamma = 1.5$ (in blue) starts to diverge from that for $\gamma = \infty$ at around $t = 1.5$; the curve for $\gamma = 1$ (in green) starts to diverge from that for $\gamma = \infty$ at around $t = 1$. Figures of various $\gamma$ values for the other three estimators (DE.LN, KM.S, KM.C) are included in Appendix \ref{appen:gamma}. Although the exact same pattern (the curve for $\gamma = \gamma_0$ diverging from that for $\gamma = \infty$ at around $t=\gamma_0$) is not observed for the other estimators, the general conclusion is that the more restrictive the selection criteria is (i.e., smaller $\gamma$), the more different the estimate is from that when there is no restriction. This is because for a given $\gamma < \infty$, no double-samples would have been enrolled in the study for longer than $\gamma$ years, and therefore little information is available regarding the beyond-$\gamma$ survival probability for the dropouts and such estimate will be largely based on extrapolation. The estimates and the confidence intervals for $t=0.5$, $1$, $1.5$, $2$ using pseudo data sets with $\gamma = \infty$, $2$, $1.5$, $1$ are listed in Table \ref{tab:data-analysis-mortality-4est}.

\section{Discussion}
\label{sec:discussion}

We proposed a deductive method to produce semiparametric estimators for estimating survival probability in the double-sampling design. The method generalizes the approach in \citet{frangakis2015deductive} by incorporating the discretized support structure and is easily computerizable. Two implementations of the method are described: one using Cox models as working models, and the other using log-normal regressions as working models. We applied the method to estimating mortality rate using data from a double-sampling component at a site of the PEPFAR program, and evaluated the impact of double-sampling selection criteria on the estimated mortality rate.

The simulation results show that the confidence interval calculated from normal approximation and the influence function-based standard error, $n^{-1}\big[ \sum_i \gateaux\{\data_i,$ $\hat{F}(\hat{\alpha}),\epsilon \}^2 \big]^{1/2}$, can be anti-conservative for the deductive estimators with lower than nominal level coverage probability, especially for small sample size such as $n=50$ or $100$. A potential alternative is to use bootstrap to obtain the confidence intervals; however, we were not able to numerically validate this due to the computational cost of bootstrapping the deductive estimator.

It would be interesting to compare the proposed estimators with estimators derived if one had used the explicit form of the EIF. While \citet{Robins2001} describe briefly the conditions that the EIF satisfies through a set of equations, we have found that by solving those equations by hand, we had high risk of introducing possible errors. We are also unaware of any work that gives a closed form expression of the EIF.




In the literature, discussions on robust estimators have been partly based on characterizing robust estimating functions; see, for example, \citet{Robins2000} and \citet{Robins2001}. Although our estimators are obtained by (numerically) solving the EIF equation, it is more difficult to find analytically all the conditions for which the estimators are consistent, precisely because the focus is on problems in which the EIF is difficult to derive analytically. Perhaps, therefore, a supplemental numerical method may exist that can also characterize more intuitively these conditions.

\textsf{R} code for the simulation and the data analysis in the paper can be downloaded at \url{https://github.com/tqian/dce_ds}.

\section*{Acknowledgments}

The authors would like to thank Ming-Wen An of Vassar College for providing the code for the estimator in \citet{an2014statmed}. The authors would also like to acknowledge Beverly S. Musick of Indiana University for compiling the database on which this study was based and for offering expert advice on the data. This work was supported by the U.S. National Institute of Drug Abuse (R01 AI102710-01). The statements in this work are solely the responsibility of the authors and do not represent the views of this organization.

\bibliographystyle{spbasic}      
\bibliography{eif-refs}

\newpage

\appendix


\section*{Appendix}

\numberwithin{equation}{section}
\numberwithin{table}{section}
\numberwithin{figure}{section}


\section{Robustness of the deductive estimator}
\label{appen:robust}

Suppose the working model assumes the true distribution $F_0$ belongs in some set $\cF$. Then the estimator, say $\hat{\tau}$, that solves the nonparametric EIF within the working model will, in the limit, be
\begin{equation}
\tau(F^*), \mbox{ for $F^* \in \cF$ that} \mbox{ solves }E_0 \phi (D_i, F^*) =0, 
\end{equation}
where $E_0$ denotes the expectation under $F_0$.
Therefore, by denoting $\bar{\phi}(F_0, F):=E_0 \phi (D_i, F)$, and assuming sufficient smoothness of the distributions, the estimator $\hat{\tau}$ will converge to the true value $\tau(F_0)$ under the following joint conditions:
\begin{equation}
\begin{cases}
 \tau(F^*)=\tau(F_0) & \mbox{ (providing correct estimand)}\\
  F^*  \mbox{ solves }\bar{\phi}(F_0, F^*)=0 & \mbox{ (fitting the model $\{F\}$ using $\phi$) }\label{condition}
\end{cases}
\end{equation}
The analytic form of above expressions may not be easily accessible when the form of the EIF is not. Suppose however, a computational method can easily determine just the "zeros" of the expressions, that is, given any $F_0$, the features of $F^*$ that are restricted in order to satisfy conditions \eqref{condition}. With such a method, coupled with the method of deductive estimation, the researcher can focus efforts to clarify and model especially well those restricted features, as this would provide approximate consistency of the estimator.


\section{Identification of $P(T > t)$}
\label{appen:identification}

Suppose $\surv_t(\cdot)$ is a function that takes an arbitrary distribution $\cL$ of $(X, \Delta)$ from independent survival and censoring times and returns the survival probability beyond $t$, $P(T > t)$ (this function is the probability limit of the Kaplan-Meier estimator \citep{kaplan1958nonparametric}). By Assumption \ref{assumpt:independent-censoring} ($T \ci C \mid Z$), we have $P(T > t \mid Z = z) = \surv_t\{ \cL(X, \Delta \mid z) \}$, where $\cL(X, \Delta \mid z)$ denotes the conditional distribution of $(X, \Delta)$ given $Z = z$. Thus,
\begin{align}
    P(T > t) = E_Z [\surv_t\{ \cL(X, \Delta \mid Z) \}]. \nonumber
\end{align}
Let $p_{X, \Delta \mid Z}(x, \delta \mid z)$ denote the probability density function of $\cL(X, \Delta \mid Z)$. We have
\begin{align}
    p_{X, \Delta \mid Z}(x, \delta \mid z) = p_{X, \Delta, \Robs \mid Z}(x, \delta, 1 \mid z) + p_{X, \Delta, \Robs \mid Z}(x, \delta, 0 \mid z). \label{eq:iden-proofuse-1}
\end{align}
By basic probability manipulation, we have
\begin{align}
    p_{X, \Delta, \Robs \mid Z}(x, \delta, 1 \mid z) &= p_{X, \Delta, \Robs, Z}(x, \delta, 1, z) / p_Z(z) \nonumber \\
    &= p_{\Robs}(1) p_{Z \mid \Robs}(z \mid 1) p_{X,\Delta \mid \Robs, Z}(x, \delta \mid 1, z) / p_Z(z), \label{eq:iden-proofuse-2}
\end{align}
and
\begin{align}
 & p_{X, \Delta, \Robs \mid Z}(x, \delta, 0 \mid z) \nonumber \\
    &= p_{X, \Delta, \Robs, Z}(x, \delta, 0, z) / p_Z(z) \nonumber \\
    &= \int p_{X, \Delta, \Robs, Z, W}(x, \delta, 0, z, w) \mu(dw) / p_Z(z) \nonumber \\
    &= \int p_{\Robs}(0) p_{Z, W \mid \Robs}(z, w \mid 0) p_{X, \Delta \mid \Robs, Z, W, S}(x, \delta \mid 0, z, w, 1) \mu(dw) / p_Z(z), \label{eq:iden-proofuse-3}
\end{align}
where $\mu$ is the dominating measure, and \eqref{eq:iden-proofuse-3} follows from Assumption \ref{assumpt:double-sampling}; i.e.,
\begin{align}
    p_{X, \Delta \mid \Robs, Z, W}(x, \delta \mid 0, z, w) = p_{X, \Delta \mid \Robs, Z, W, S}(x, \delta \mid 0, z, w, 1). \nonumber    
\end{align}
Plugging \eqref{eq:iden-proofuse-2} and \eqref{eq:iden-proofuse-3} into \eqref{eq:iden-proofuse-1}, we have that $p_{X, \Delta \mid Z}(x, \delta \mid z)$, and hence $P(T > t)$, is identifiable from the observed data distribution components \eqref{eq:distr1} and \eqref{eq:distr2}.


\section{Computing \eqref{eq:distr1} and \eqref{eq:distr2} from an arbitrary distribution on $\Omega$}
\label{appen:averaging-normalization}

Suppose $G$ is an arbitrary probability distribution on $\Omega = \Omega_0 \cup \Omega_1$. Because $\Omega$ is discrete and so is $G$, we will use $P(\cdot)$ instead $\pr(\cdot)$. For an arbitrary point $o \in \Omega$, we use $o(z)$ to denote the entry of $o$ corresponding to $Z$, and similarly for $o(x)$, $o(\delta)$, etc. We define $0/0 = 0$. The distribution components \eqref{eq:distr1} and \eqref{eq:distr2} can be computed as follows.

\begin{align}
    P(\Robs = 1) & = \sum_{o\in\Omega_1} G(o), \nonumber \\
    P(Z = z^* \mid \Robs = 1) & = \frac{\sum_{o \in \Omega_1} G(o) \indic\{o(z) = z^*\}}{\sum_{o \in \Omega_1} G(o)}, \nonumber \\
    P(X = x^*, \Delta = \delta^* \mid Z = z^*, \Robs = 1) & = \frac{\sum_{o \in \Omega_1} G(o) \indic\{o(x) = x^*, o(\delta) = \delta^*, o(z) = z^*\} }{\sum_{o \in \Omega_1} G(o) \indic\{o(z) = z^*\}}, \nonumber \\
    P(\Robs = 0) & = 1 - P(\Robs = 1), \nonumber \\
    P(Z = z^*, W = w^* \mid \Robs = 0) & = \frac{\sum_{o \in \Omega_0} G(o) \indic\{o(z) = z^*, o(w) = w^*\}}{\sum_{o \in \Omega_0} G(o)}, \nonumber
\end{align}
and
\begin{align}
    & P(X = x^*, \Delta = \delta^* \mid Z = z^*, W = w^*, \Robs = 0, S = 1) \nonumber \\
    & = \frac{\sum_{o \in \Omega_0} G(o) \indic\{o(x) = x^*, o(\delta) = \delta^*, o(z) = z^*, o(w) = w^*, s = 1\} }{\sum_{o \in \Omega_1} G(o) \indic\{o(z) = z^*, o(w) = w^*, s = 1\}}. \nonumber
\end{align}

\section{Details about the computation of terms in \eqref{eq:calculate-hatF}}
\label{appen:working-distributions}

Recall that the observed data for patient $i$ is denoted by $(\robs_i, z_i, w_i, s_i, x_i, \delta_i)$, and we assume that the data is sorted so that $\robs_i = 0$ for $1\leq i \leq m$, $\robs_i = 1$ for $m+1 \leq i \leq n$, $s_i = 1$ for $1 \leq i \leq m_1$, and $s_i = 0$ for $m_1 + 1 \leq i \leq n$, and that $x_1 < x_2 < \ldots < x_{m_1}$ and $x_{m+1} < x_{m+2} < \ldots < x_n$. This last assumption that no ties in $x$ values is for notation simplicity in the following exposition; our estimation program allows ties in $x$ values.

Let $o = (\robs, z, w, s, x, \delta)$ denote an arbitrary point in $\Omega$ (see Section \ref{subsec:algorithm} for the definition of $\Omega$), and let $P_{\hat{F}}(o)$ denote the probability of $o$ under the distribution $\hat{F}$. We rewrite \eqref{eq:calculate-hatF} here for reference:
\begin{align}
    P_{\hat{F}}(\robs, z, w, s, x, \delta) = & \hat{P}(\Robs = \robs) \hat{P}(Z=z, W=w \mid \Robs = \robs) \nonumber \\
        & \times \hat{P}(S=s \mid \Robs = \robs, Z=z, W=w) \nonumber \\
        & \times \hat{P}(X=x, \Delta=\delta \mid \Robs = \robs, Z=z, W=w, S=s). \nonumber
\end{align}
Each of the terms on the right hand side is computed as follows:

\begin{itemize}
    \item $\hat{P}(\Robs = \robs)$: the probability of being an observed dropout. This is fitted by the empirical frequency: $\hat{P}(\Robs = \robs) = n^{-1} \sum_{i=1}^n \indic(\robs_i = \robs)$.
    
    \item $\hat{P}(Z=z, W=w \mid \Robs = \robs)$. This is fitted by the empirical frequency: 
    \begin{align*}
        \hat{P}(Z=z, W=w \mid \Robs = \robs) = \frac{\sum_{i=1}^n \indic(\robs_i = \robs, z_i=z, w_i=w)}{\sum_{i=1}^n \indic(\robs_i = \robs)}.
    \end{align*}
    
    \item $\hat{P}(S=1 \mid \Robs = 0, Z=z, W=w)$. This is fitted by logistic regression using data from patients with $\robs_i = 0$ and with $Z,W$ being regressors. For $\Robs = 1$, $S$ is always equal to 0 by the construction of $\Omega$.
    
    \item $\hat{P}(X = x, \Delta = \delta \mid \Robs = \robs, Z=z, W=w, S = s)$. For $(\robs = 0, s=0)$, by the construction of $\Omega$ we have $(x,\delta) = (\na, \na)$, and hence the corresponding probability equals 1. For $(\robs = 0, s=1)$ or $\robs = 1$, this probability is fitted using the likelihood arising from one of two working models: Cox proportional hazards model \citep{cox1972regression} or log-normal regression. In both working models, the working assumption of conditional independent censoring ($T \ci C \mid \Robs, Z, W, S$) is used.
    \begin{itemize}
        \item When Cox proportional hazards model is used as the working model, we fit four separate Cox models without sharing parameters across models:
        \begin{align}
            & T \sim Z, W \mid \Robs = 0, S = 1; && C \sim Z, W \mid \Robs = 0, S = 1; \nonumber \\
            & T \sim Z, W \mid \Robs = 1; && C \sim Z, W \mid \Robs = 1. \label{appen:eq:tc-model-fits}
        \end{align}
        For example, $T \sim Z, W \mid \Robs = 0, S = 1$ means that we fit a Cox model with $T$ being the failure time (hence $C$ being the censoring time), with $Z,W$ as regressors, and using data from patients with $\Robs = 0, S=1$; $C \sim Z, W \mid \Robs = 1$ means that we fit a Cox model with $C$ being the failure time (hence $T$ being the censoring time), with $Z,W$ as regressors, and using data from patients with $\Robs = 1$. For each model the baseline hazard is estimated using Breslow's method \citep{breslow1974covariance}. This gives the following survival probabilities:
        \begin{align}
            S_T^{(0)}(x; z,w) & := \hat{P}(T > x \mid Z = z, W = w, \Robs = 0, S=1) \nonumber \\
            S_C^{(0)}(x; z,w) & := \hat{P}(C > x \mid Z = z, W = w, \Robs = 0, S=1) \nonumber
        \end{align}
        for $x \in \{ x_1, \ldots, x_{m_1}\}$ (because here $(x,z,w)\in\Omega_0$\footnote{We are abusing the notation slightly; more precisely, we meant $(\robs =0,z,w, s=1, x, \delta)\in\Omega_0$.}), and 
        \begin{align}
            S_T^{(1)}(x; z, w) & := \hat{P}(T > x \mid Z = z, W = w, \Robs = 1) \nonumber \\
            S_C^{(1)}(x; z, w) & := \hat{P}(C > x \mid Z = z, W = w, \Robs = 1) \nonumber
        \end{align}
        for $x \in \{ x_{m+1}, \ldots, x_n\}$ (because here $(x,z,w)\in\Omega_1$). The survival probabilities (cumulative) are then converted to discrete probability mass as follows. For $(x,z,w)\in\Omega_0$, suppose $x=x_i$ for some $1 \leq i \leq m_1$, then
        \begin{align}
            p_T^{(0)}(x; z,w) & := \hat{P}(T = x \mid Z = z, W = w, \Robs = 0, S=1) \nonumber \\
            & = \begin{cases}
                    1 - S_T^{(0)}(x_1;z,w) & \mbox{if } i = 1,\\
                    S_T^{(0)}(x_{i-1};z,w) - S_T^{(0)}(x_i;z,w) & \mbox{otherwise}.
                \end{cases} \nonumber
        \end{align}
        Similarly, we calculate
        \begin{align}
            p_C^{(0)}(x; z,w) := \hat{P}(C = x \mid Z = z, W = w, \Robs = 0, S=1). \nonumber
        \end{align}
        The probability on $(X,\Delta)$ for $\Omega_0$ is calculated as
        \begin{align}
            & \hat{P}(X = x, \Delta = \delta \mid \Robs = 0, Z = z, W = w, S = 1) \nonumber \\
            & = \{p_T^{(0)}(x;z,w) S_C^{(0)}(x;z,w)\}^\delta ~ \{S_T^{(0)}(x;z,w) p_C^{(0)}(x;z,w)\}^{(1-\delta)}, \label{appen:eq:prob-xdelta}
        \end{align}
        and is further normalized over $x \in \{x_1, \ldots, x_{m_1}\}$ (within each $(z,w)$ pair) to ensure
        \begin{align}
            \sum_{i=1}^{m_1} \hat{P}(X = x_i, \Delta = \delta_i \mid \Robs = 0, Z=z, W=w, S=1) = 1. \label{appen:eq:prob-xdelta-normalization}
        \end{align}
        $p_T^{(1)}(x; z,w)$, $p_C^{(1)}(x; z,w)$, and the probability on $(X,\Delta)$ for $\Omega_1$, $\hat{P}(X = x, \Delta = \delta \mid \Robs = 1, Z = z, W = w)$, is calculated similarly.
        
        \item When log-normal regression is used as the working model, we fit the four log-normal regressions \eqref{appen:eq:tc-model-fits} using \textsf{survreg} function in \textsf{R} package \textsf{survival} \citep{Rsurvival}, without sharing parameters across models. This gives the following probability density functions (with respect to Lebesgue measure):
        \begin{align}
            f_T^{(0)}(t;z,w) & := p_{ T\mid Z,W, \Robs = 0, S = 1}(t \mid z, w), \nonumber \\
            f_C^{(0)}(t;z,w) & := p_{ C\mid Z,W, \Robs = 0, S = 1}(t \mid z, w), \nonumber \\
            f_T^{(1)}(t;z,w) & := p_{ T\mid Z,W, \Robs = 1}(t \mid z, w), \nonumber \\
            f_C^{(1)}(t;z,w) & := p_{ C\mid Z,W, \Robs = 1}(t \mid z, w). \nonumber
        \end{align}
        These probability density functions are then converted to probability mass functions with positive probability only at points in $\Omega_0$ and $\Omega_1$, respectively. For example, for $(x,z,w)\in\Omega_0$, suppose $x=x_i$ for some $1 \leq i \leq m_1$, then
        \begin{align}
            p_T^{(0)}(x; z,w) & := \hat{P}(T = x \mid Z = z, W = w, \Robs = 0, S=1) \nonumber \\
            & = \begin{cases}
                    \int_0^{(x_1 + x_2)/2} \, f_T^{(0)}(t;z,w) \, dt & \mbox{if } i = 1,\\
                    \int_{(x_{i-1} + x_i)/2}^{(x_i + x_{i+1})/2} \, f_T^{(0)}(t;z,w) \, dt & \mbox{if } 2 \leq i \leq m_1 - 1, \\
                    \int_{(x_{m_1-1} + x_{m_1})/2}^{\infty} \, f_T^{(0)}(t;z,w) \, dt & \mbox{if } i = m_1. \\
                \end{cases} \nonumber
        \end{align}
        $p_C^{(0)}(x; z,w)$, $p_T^{(1)}(x; z,w)$ and $p_T^{(1)}(x; z,w)$ are computed similarly. The survival probabilities (such as $S_T^{(0)}(x; z,w)$) are then calculated from the probability mass functions, and the probability on $(X,\Delta)$ is calculated as \eqref{appen:eq:prob-xdelta} and further normalized to ensure \eqref{appen:eq:prob-xdelta-normalization}.
    \end{itemize}
    
\end{itemize}

\section{Additional results regarding the impact of $\gamma$ on the estimated mortality rate}
\label{appen:gamma}

\begin{figure}[htbp]
    \centering
    \includegraphics[width = 0.48\textwidth]{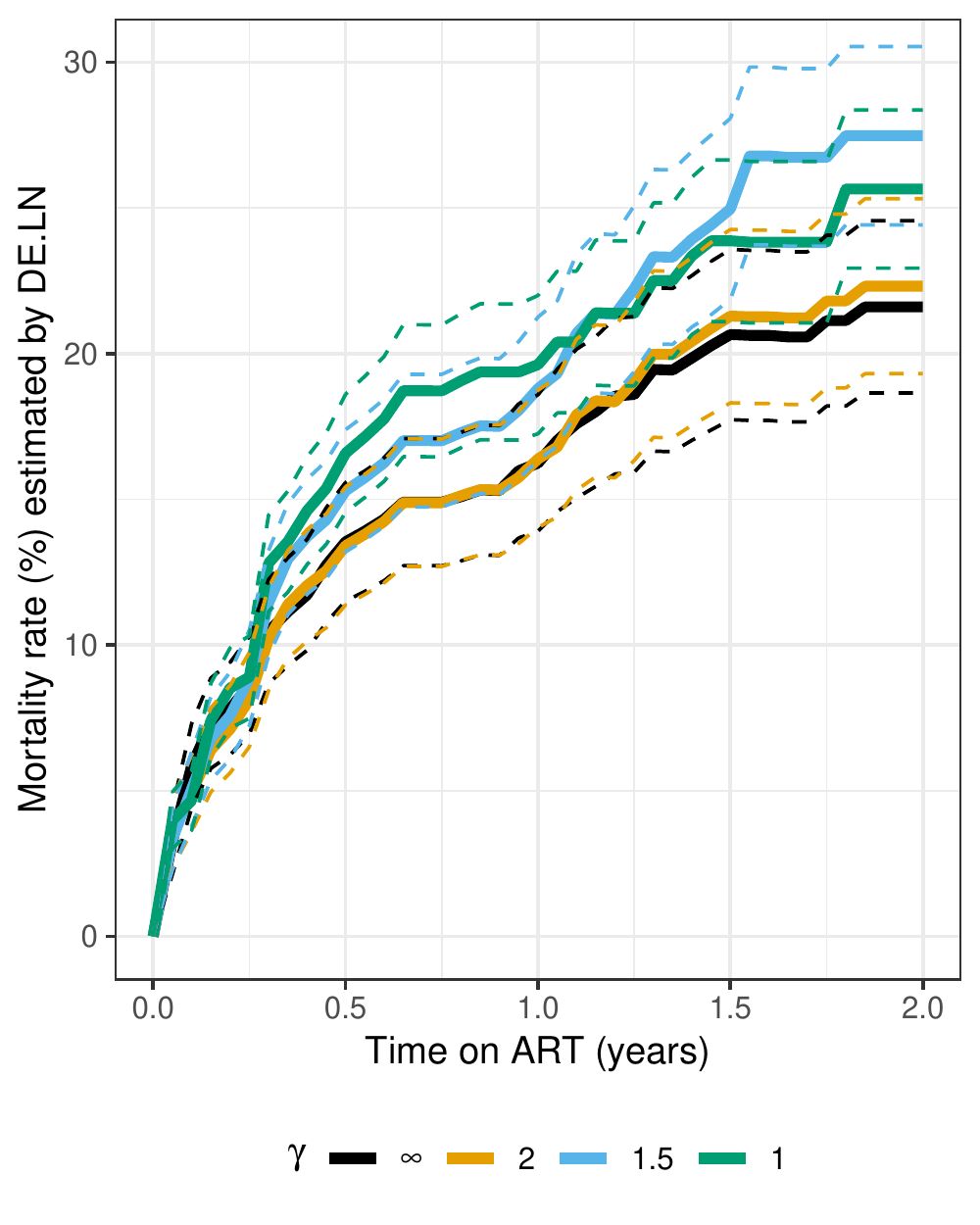}
    \caption{Estimated mortality rate (solid line) with 95\% pointwise confidence interval (dashed line) estimated by DE.LN, using PEPFAR patients data with various $\gamma$ restrictions: all double-samples included ($\gamma = \infty$), past 2-year dropouts included ($\gamma = 2$), past 1.5-year dropouts included ($\gamma = 1.5$), past 1-year dropouts included ($\gamma = 1$).}
    \label{appen:fig:DE.LN}
\end{figure}

\begin{figure}[htbp]
    \centering
    \includegraphics[width = 0.48\textwidth]{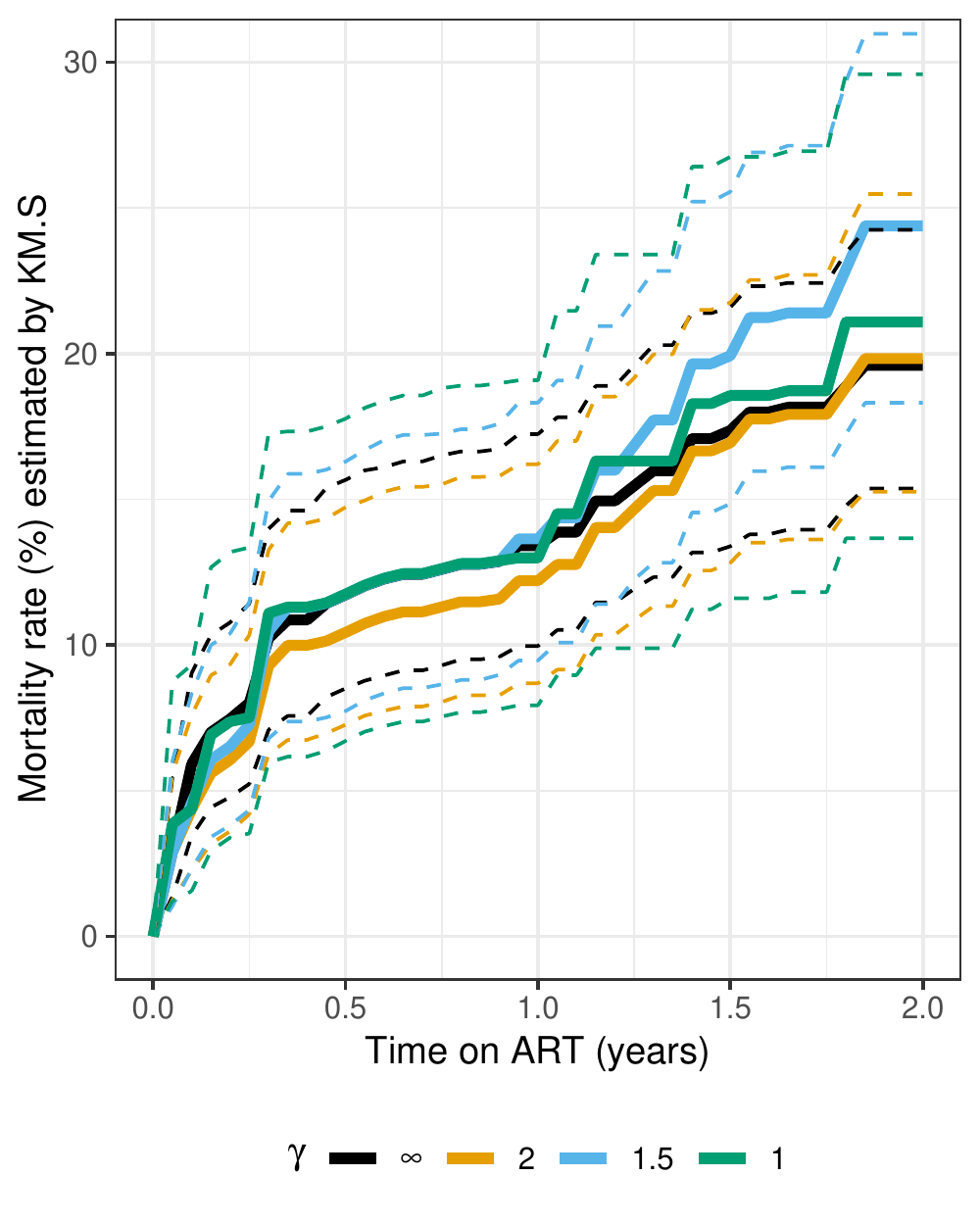}
    \caption{Estimated mortality rate (solid line) with 95\% pointwise confidence interval (dashed line) estimated by KM.S, using PEPFAR patients data with various $\gamma$ restrictions: all double-samples included ($\gamma = \infty$), past 2-year dropouts included ($\gamma = 2$), past 1.5-year dropouts included ($\gamma = 1.5$), past 1-year dropouts included ($\gamma = 1$).}
    \label{appen:fig:KM.S}
\end{figure}

\begin{figure}[htbp]
    \centering
    \includegraphics[width = 0.48\textwidth]{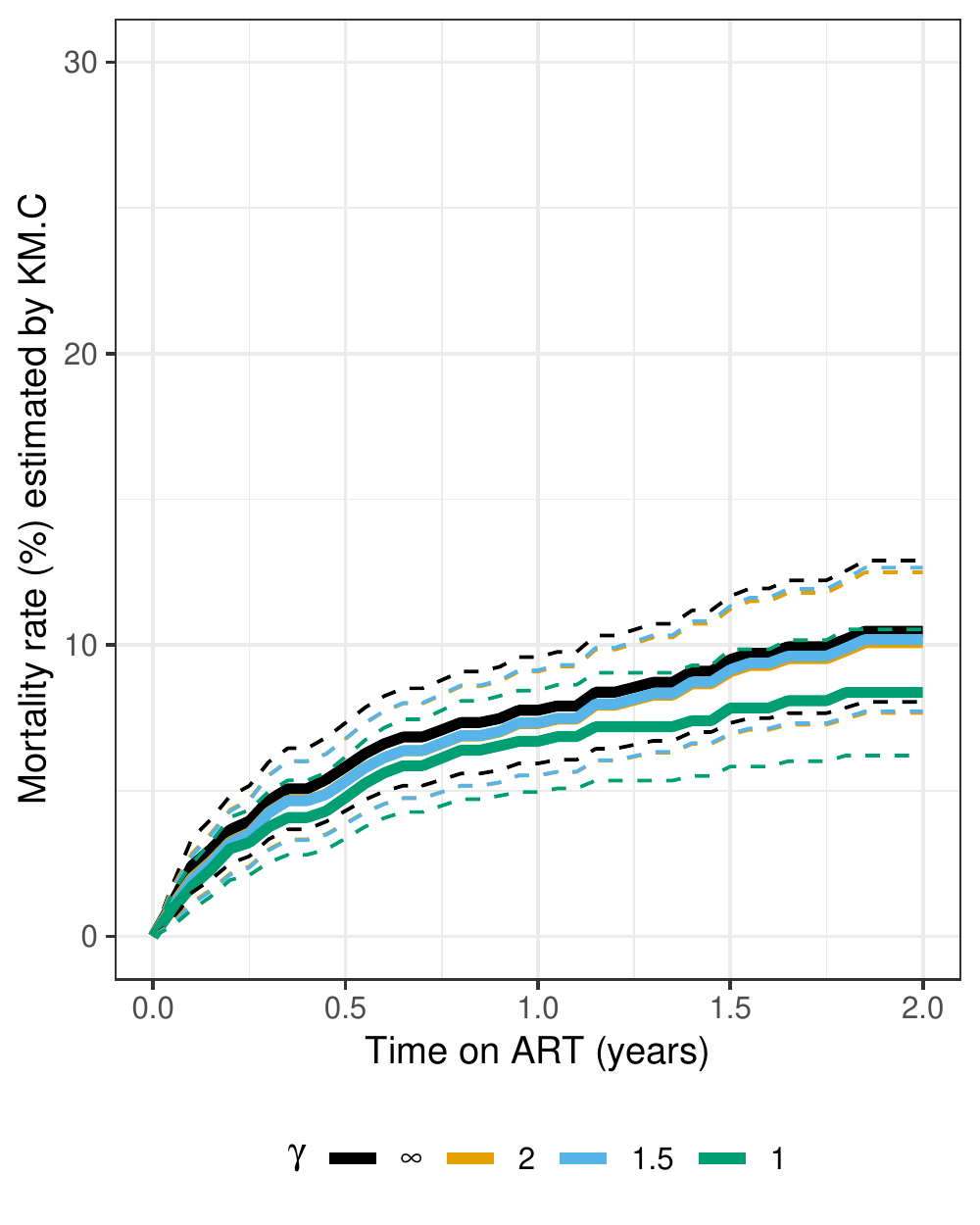}
    \caption{Estimated mortality rate (solid line) with 95\% pointwise confidence interval (dashed line) estimated by KM.C, using PEPFAR patients data with various $\gamma$ restrictions: all double-samples included ($\gamma = \infty$), past 2-year dropouts included ($\gamma = 2$), past 1.5-year dropouts included ($\gamma = 1.5$), past 1-year dropouts included ($\gamma = 1$).}
    \label{appen:fig:KM.C}
\end{figure}

Here we include results of the estimated mortality rate based on pseudo data sets with various $\gamma$ values, using three estimators (DE.LN in Figure \ref{appen:fig:DE.LN}, KM.S in Figure \ref{appen:fig:KM.S}, KM.C in Figure \ref{appen:fig:KM.C}). The result of the estimator DE.Cox has been presented and discussed in the ``\textit{Impact of the double-sampling selection criteria on the estimated mortality rate}'' paragraph of Section \ref{sec:application}.

\label{lastpage}

\end{document}